\documentclass[a4paper,11pt]{article}
\pdfoutput=1
\usepackage{jheppub}

\usepackage{amssymb,amsmath,amsfonts}
\usepackage[normalem]{ulem}
\usepackage[utf8x]{inputenc}
\usepackage{slashed}
\usepackage{graphicx}
\usepackage{tabularx}
\usepackage{here}
\usepackage{color}
\usepackage{csquotes} 
\usepackage{comment}
\usepackage{mathrsfs}
\usepackage{float}
\usepackage{ascmac}
\usepackage{multirow}
\usepackage{longtable}
\usepackage{bm}
\usepackage{ulem}
\usepackage{caption}
\usepackage{nicematrix}
\usepackage{arydshln}

\usepackage[italicdiff]{physics}
\usepackage{tikz}
\usepackage[compat=1.1.0]{tikz-feynhand}
\makeatletter
\newcommand*\rel@kern[1]{\kern#1\dimexpr\macc@kerna}
\newcommand*\widebar[1]{%
  \begingroup
  \def\mathaccent##1##2{%
    \rel@kern{0.8}%
    \overline{\rel@kern{-0.8}\macc@nucleus\rel@kern{0.2}}%
    \rel@kern{-0.2}%
  }%
  \macc@depth\@ne
  \let\math@bgroup\@empty \let\math@egroup\macc@set@skewchar
  \mathsurround\z@ \frozen@everymath{\mathgroup\macc@group\relax}%
  \macc@set@skewchar\relax
  \let\mathaccentV\macc@nested@a
  \macc@nested@a\relax111{#1}%
  \endgroup
}
\makeatother

\numberwithin{equation}{section}

\preprint{
\begin{minipage}{5cm}
\small
\flushright
EPHOU-26-10\\ 
KYUSHU-HET-370
\end{minipage}}

\title{
Note on arithmetic structure of modulus vacua in flux compactifications
}

\author{Yukiya Furuta$^1$,} 
\author{Tatsuo Kobayashi$^1$,} 
\author{Tomoyasu Kori$^1$,}
\author{Shuhei Miyamoto$^1$,} 
\author{Ryusei Nishida$^1$, and } 
\author{Hajime Otsuka$^{2,3}$}
\affiliation{
$^1$Department of Physics, Hokkaido University, Sapporo 060-0810, Japan\\
$^2$Department of Physics, Kyushu University, 744 Motooka, Nishi-ku, Fukuoka 819-0395, Japan
\\
$^3$Quantum and Spacetime Research Institute (QuaSR), Kyushu University, 744 Motooka, Nishi-ku, Fukuoka 819-0395, Japan
}

\emailAdd{y-furuta@particle.sci.hokudai.ac.jp}
\emailAdd{kobayashi@particle.sci.hokudai.ac.jp}
\emailAdd{t-kori@particle.sci.hokudai.ac.jp}
\emailAdd{s-miyamoto@particle.sci.hokudai.ac.jp}
\emailAdd{r-nishida@particle.sci.hokudai.ac.jp}
\emailAdd{otsuka.hajime@phys.kyushu-u.ac.jp}

\abstract{We study modulus stabilization by background fluxes. 
{Within the truncated complex-structure sector, the F-term conditions of the complex structure moduli are described by a holomorphic quadratic equation.} 
As concrete examples, we consider {the} $T^6/(\mathbb{Z}_2\times\mathbb{Z}_2')$ orientifold model and a simple Calabi-Yau compactification. 
The modulus values show specific patterns. 
For example, they {exhibit a structure related to} the
Farey sequence.
The void structure appears around modulus vacua with high degeneracies.
We find a correlation between the degeneracy and the void area. 
The modulus vacua are related by discrete Abelian symmetries generated by the Gauss composition law, 
which includes the CP symmetry.
Spontaneous CP violation is also discussed.
}

\makeatletter
\gdef\@fpheader{}
\makeatother

\begin{document}

\maketitle

\section{Introduction}
\label{sec:intro}

Moduli stabilization is an important issue in four-dimensional low energy effective field theory derived from string theory.
{Background Ramond-Ramond (RR) and Neveu-Schwarz (NS) fluxes provide a well-established mechanism for moduli stabilisation} \cite{Giddings:2001yu}.\footnote{Also, see e.g. Refs.~\cite{Grana:2000jj,Dasgupta:1999ss,Becker:1996gj}.}
See for review Refs.~\cite{Grana:2005jc,Douglas:2006es,McAllister:2025qwq}.
3-form fluxes induce superpotential terms $W$ including mass terms of some moduli such as complex structure moduli  \cite{Gukov:1999ya}, and these moduli are stabilized at supersymmetric vacua.
Furthermore, non-perturbative effects and quantum effects can stabilize other moduli, which remain light by flux compactification \cite{Kachru:2003aw,Becker:2002nn,Balasubramanian:2004uy,Balasubramanian:2005zx,McAllister:2024lnt}.
Supersymmetry (SUSY) may be broken at this stage.

Moduli stabilization is important in particle physics as well as cosmology. 
All couplings such as gauge couplings and Yukawa couplings are functions of moduli \cite{Ibanez:2012zz}.
That is, moduli values determine these coupling values.
For example, {in modular flavor scenarios}, Yukawa couplings are given by modular forms.\footnote{Indeed, modular flavor symmetries and modular forms were studied in heterotic orbifold models \cite{Ferrara:1989qb,Lerche:1989cs,Lauer:1989ax,Lauer:1990tm} and magnetized compactification of type IIB theory \cite{Kobayashi:2018rad,Kobayashi:2018bff,Ohki:2020bpo,Kikuchi:2020frp,Kikuchi:2020nxn,
Kikuchi:2021ogn,Almumin:2021fbk}.}
(See Refs.~\cite{Feruglio:2017spp,Kobayashi:2018vbk,Penedo:2018nmg,Criado:2018thu,Kobayashi:2018scp,Novichkov:2018ovf,Novichkov:2018nkm,deAnda:2018ecu,Okada:2018yrn,Kobayashi:2018wkl,Novichkov:2018yse} for earlier works and Refs.~\cite{Kobayashi:2023zzc,Ding:2023htn} for reviews.)
Thus, moduli values are important to derive realistic quark and lepton masses and their mixing angles as well as CP phases.

As mentioned above, the moduli values are also relevant to CP phases.
String theory is CP symmetric \cite{Green:1987mn,Strominger:1985ku}.
That is good for the strong CP phase \cite{Dine:1992ya,Choi:1992xp}, but we cannot realize the weak CP phase.
The modulus $u$ transforms as $u \to -\bar u$ under the CP transformation in four-dimensional low energy effective field theory. 
One scenario is that CP symmetry may be violated explicitly by background through compactification.
Indeed, {a generic flux background violates CP symmetry}, $u \longleftrightarrow -\bar u$. 
An alternative interesting scenario is that background still preserves CP symmetry, but it may be violated spontaneously at the potential minimum.
Indeed, spontaneous CP violation was studied in Refs.~\cite{Kobayashi:2020uaj,Ishiguro:2020nuf}, where $|W|^2$ is invariant under the CP transformation and  flux background still preserves CP symmetry.

Distribution of modulus values stabilized by flux compactification, i.e., flux vacua, was also studied \cite{Ashok:2003gk,Douglas:2003um,Denef:2004ze,DeWolfe:2004ns,Cole:2019enn,Cole:2021nnt,Dubey:2023dvu,Ebelt:2023clh,Chauhan:2025rdj}.
Fluxes themselves must be quantized.
Flux vacua are discrete.
They show specific patterns in moduli spaces.
Such analysis on distribution of flux vacua was also applied to modular flavor models in Ref.~\cite{Ishiguro:2020tmo}, which can show statistically favorable moduli values in modular flavor models. 
In addition to such statistical distributions, it is also interesting to understand analytically the arithmetic structure of the modulus values, the degeneracy of flux configurations leading to the same vacuum, and relations among different vacua.

We revisit modulus stabilization by background fluxes.
Rather, we focus on a simple model such as a single complex structure modulus within the framework of type IIB theory.
A similar model can be obtained in heterotic string theory and type IIA theory. 
We investigate distribution of modulus vacua. 
By discussing a simple model, we can carry out detailed study of the distribution of vacua and their degeneracy factors by different combinations of fluxes analytically, systematically and concretely. 
In the explicit type IIB models, we introduce RR fluxes and concentrate on {solutions of the F-term conditions in the truncated complex-structure sector.} We show that the SUSY condition for a single modulus can be reduced to a holomorphic quadratic equation, whose coefficients are determined by fluxes. This quadratic equation is useful for analyzing the distribution and relations of modulus vacua. 
As concrete examples, we consider the $T^6/(\mathbb{Z}_2\times\mathbb{Z}_2)$ orientifold model and a simple Calabi-Yau compactification. We find specific patterns of modulus values. In particular, for Re~$u = 0$ in the toroidal model, $(\mathrm{Im}~u)^2$ is related to the Farey sequence. We also study the degeneracy of flux configurations and the void structure in the distribution of vacua, and find a correlation between the degeneracy and the size of the void. Furthermore, the quadratic equations with the same discriminant are related by the Gauss composition law, which gives Abelian relations among vacua and includes the CP transformation. We also discuss spontaneous CP violation and find that it is difficult to realize it from the CP-symmetric flux superpotentials.

This paper is organized as follows.
In section \ref{sec:setup}, we explain our setup on supergravity theory with flux compactification.
In section \ref{sec:orientifol}, we 
consider the overall complex structure modulus in the $T^6/(\mathbb{Z}_2\times\mathbb{Z}_2')$ orientifold  model of type IIB theory.
We study the modulus stabilization by introducing 3-form fluxes. 
We investigate the F-term condition of complex structure moduli and show the distribution of vacua. 
We also study the degeneracy and void structure of the vacua, as well as relations among vacua by modular transformations and Gauss composition. 
We also discuss the possibility for the spontaneous CP violation.
In section \ref{sec:CY}, we extend the analysis on the $T^6/(\mathbb{Z}_2\times\mathbb{Z}_2')$ orientifold model to one in a simple Calabi-Yau compactification with a single modulus.
In section \ref{sec:multi-moduli}, we study models with multi moduli fields and discuss the possibility for the spontaneous CP violation.
Section \ref{sec:conclusion} is devoted to the conclusions.
In Appendix \ref{app:modular-form}, we discuss the modular forms, which are relevant to modulus vacua by flux compactification.

\section{Moduli stabilization}
\label{sec:setup}

Here, we explain our setup for moduli stabilization by background fluxes.
We study the stabilization of complex structure moduli $u^a$ within the framework of type IIB string theory.
{We neglect the dynamics of moduli other than $u^a$ such as the axio-dilaton $s$ and K\"ahler moduli $t^\alpha$ and focus on the stabilization of $u^a$ by the superpotential  $W(u)$ due to RR background fluxes.}

{In the full theory with moduli $s,t^\alpha, u^a$, an additional superpotential $\Delta W(s,t^\alpha)$ may be induced in addition to $W(u)$.
If $u^a$ are stabilized at sufficiently heavy masses by $W(u)$ compared with the remaining moduli and the condition $|W(u)| \gg |\Delta W(s,t^\alpha)|$ is satisfied, we expect only perturbatively small corrections to the arithmetic structure of the complex-structure solutions. This assumption does not ensure supersymmetry of the full theory. 
Alternatively, when both RR fluxes and NS fluxes are introduced, such backgrounds induce superpotential terms $W(u^a,s)$, where $u^a$ and $s$ are mixed. 
For such backgrounds, we need to extend the following analyses. 
At any rate, our purpose is to study the arithmetic structure of moduli vacua analytically, systematically, and concretely in simple models before such extensions.
Thus, we consider only the truncated sector of  $u^a$.}

{More precisely, in the full type IIB theory the Gukov-Vafa-Witten superpotential is}
\begin{align}
    {W_{\rm GVW}(u,s)=\int (F_3- s H_3)\wedge\Omega(u)
    =P_F(u)-s P_H(u).}
\end{align}
{For the RR-flux background considered here, $H_3=0$ and hence $P_H(u)=0$. The axio-dilaton F-term condition is then}
\begin{align}
    {D_s W=(\partial_s K)W=0.}
\end{align}
{Consequently, generic solutions of $D_aW=0$ with $W\neq 0$ are not supersymmetric vacua of the complete type IIB theory. We also do not impose the F-term conditions for the K\"ahler moduli. Throughout the remainder of this paper, the terms ``SUSY condition'' and ``SUSY solution'' refer only to the F-term conditions in the truncated complex-structure sector, unless otherwise stated.}
{With $H_3\neq 0$, the equations $D_aW=0$ and $D_sW=0$ must be solved simultaneously for $u^a$ and $s$. After eliminating $s$, the resulting equations for $u^a$ need not be quadratic. The RR and NS fluxes must also satisfy the D3-brane tadpole condition. We leave this coupled problem for future work.}

{The  K\"ahler potential $K$  of $u^a$} is written by 
\begin{align}
    K=-\ln (\kappa ), \qquad 
    \kappa=\frac{i}{6}\kappa_{abc} (u^a-\bar u^{\bar a})(u^b-\bar u^{\bar b})(u^c-\bar u^{\bar c} ),
\end{align}
where $\kappa_{abc}$ denotes intersection numbers.
{We work in units in which $M_{\rm P}=1$}, where $M_{\rm P}$ is the reduced Planck mass.

We introduce RR fluxes, which induce the superpotential $W$ \cite{Gukov:1999ya}. 
Note that since we do not introduce NS fluxes, RR fluxes do not contribute to the tadpole cancellation condition.
A similar supergravity Lagrangian can be realized in heterotic string theory with NS flux background, 
and also in type IIA string theory for K\"ahler moduli.
Thus, our following discussions are applicable in these heterotic string theory and type IIA string theory with flux compactifications.

{Within this truncated sector, we concentrate on solutions of the F-term conditions for the complex-structure moduli, i.e.,}
\begin{align}
    D_aW=W_a+K_aW=W_a-\frac{\kappa_a}{\kappa}W=0.
\end{align}
We require $\kappa \neq 0$.
{The complex-structure F-term condition then takes the form}
\begin{equation}\label{eq:SUSY-min}
\kappa W_a - \kappa_a W = 0.
\end{equation}
{In the next sections, we solve this equation and study the resulting distributions in the $T^6/(\mathbb{Z}_2\times\mathbb{Z}_2')$ orientifold model and in a simple Calabi-Yau model.}

\section{$T^6/(\mathbb{Z}_2\times\mathbb{Z}_2')$ orientifold model}
\label{sec:orientifol}

Here, we study moduli stabilization in $T^6/(\mathbb{Z}_2\times\mathbb{Z}_2')$ orientifold model by RR 3-form background.
We restrict ourselves to the overall complex structure modulus $u$ of $T^6/(\mathbb{Z}_2\times\mathbb{Z}_2')$.
The K\"ahler potential is written by
\begin{align}
\label{eq:K-orientifold}
    K = - \ln \kappa, \qquad \kappa \equiv i (u - \bar{u})^3,
\end{align}
and the RR fluxes induce the following superpotential:
\begin{align}
    \label{eq:W-orientifold}
 \quad W(u) = e_0 + 3e u + 3m u^2 + m_0 u^3,
\end{align}
where $e_0,e, m, m_0$ are integers.
By use of these, we study the SUSY condition (\ref{eq:SUSY-min}).
We analyze the SUSY condition by three methods in what follows.

\subsection{Gr\"obner basis}
\label{sec:grobner}

Here, we analyze the SUSY condition by use of Gr\"obner basis.\footnote{{See Ref.~\cite{Gray:2006gn} for an early work using Gr\"obner basis in the context of flux compactifications.}}
Here we decompose the modulus $u$ into its real and imaginary parts as 
\begin{align}
    u=\alpha + i\beta, \quad {\rm Re} ~u=\alpha, \quad {\rm Im}~ u=\beta . 
\end{align}
Also, we write real and imaginary parts of the SUSY condition (\ref{eq:SUSY-min}) by
\begin{equation}
h : \begin{cases}
  \kappa ({\rm Re} W_u) - i\kappa_u ({\rm Im} W) = 0, \\
  \kappa ({\rm \Im} W_u) + i\kappa_u ({\rm Re} W) = 0. \\
\end{cases}
\end{equation}
These equations generate the ideal \(I \equiv \langle h\rangle : \langle V\rangle^\infty\), and its Gr\"obner cover~\cite{Montes:2010,Montes:2018Book} is obtained as 
\begin{equation}\label{eq:orientifold_groebner}{
\begin{cases}
  4\beta^2e^2m_0^2 - 8\beta^2em^2m_0 + 4\beta^2m^4 + e_0^2m_0^2 - 6e_0emm_0 + 4e_0m^3 + 4e^3m_0 - 3e^2m^2=0, \\
  2\alpha em_0 - 2\alpha m^2 + e_0m_0 - em=0, \\
  \alpha e_0m_0 - \alpha em - 2\beta^2em_0 + 2\beta^2m^2 + 2e_0m - 2e^2=0, \\
  2\alpha e_0m^2 - 2\alpha e^2m - 4\beta^2e^2m_0 + 4\beta^2em^2 - e_0^2m_0 + 5e_0em - 4e^3=0, \\
  \alpha^2m_0 + 2\alpha m + \beta^2m_0 + e=0, \\
  \alpha^2m + 2\alpha e + \beta^2m + e_0=0, \\
\end{cases}
}\end{equation} 
under the condition 
\(2m^2-2em_0 \neq 0\),
which we computed using the \texttt{grobcov.lib} library~\cite{Montes:GrobcovLib} for the \textsc{Singular}~\cite{DGPS} computer algebra system.

The first and second equations mean that the solution $u=\alpha + i\beta$ of the SUSY condition is written by 
\begin{equation}\label{eq:orientifold_alpha}{
\alpha = \frac{-(e_0m_0-em)}{2(em_0-m^2)},
}\end{equation}
\begin{equation}\label{eq:orientifold_beta}{
\beta = \sqrt{\frac{4(e_0m-e^2)(em_0-m^2)-(e_0m_0-em)^2}{4(em_0-m^2)^2}}.
}\end{equation} 
The third and fourth equations are satisfied if the above equations (\ref{eq:orientifold_alpha}) and (\ref{eq:orientifold_beta}) are satisfied.

The fifth and sixth equations in (\ref{eq:orientifold_groebner}) are written by 
\begin{equation}\label{eq:orientifold_circ_m_m0}
\left( \alpha + \frac{m}{m_0}\right)^2 + \beta^2 = -\frac{em_0-m^2}{m_0^2}, 
\end{equation}
\begin{equation}\label{eq:orientifold_circ_e_m}
\left(\alpha + \frac{e}{m}\right)^2 + \beta^2 = -\frac{e_0m-e^2}{m^2}.
\end{equation} 
These represent circles.
These equations are also satisfied if the above equations (\ref{eq:orientifold_alpha}) and (\ref{eq:orientifold_beta}) are satisfied.

\begin{figure}
\centering
\includegraphics[width=0.6\linewidth,height=\textheight,keepaspectratio]{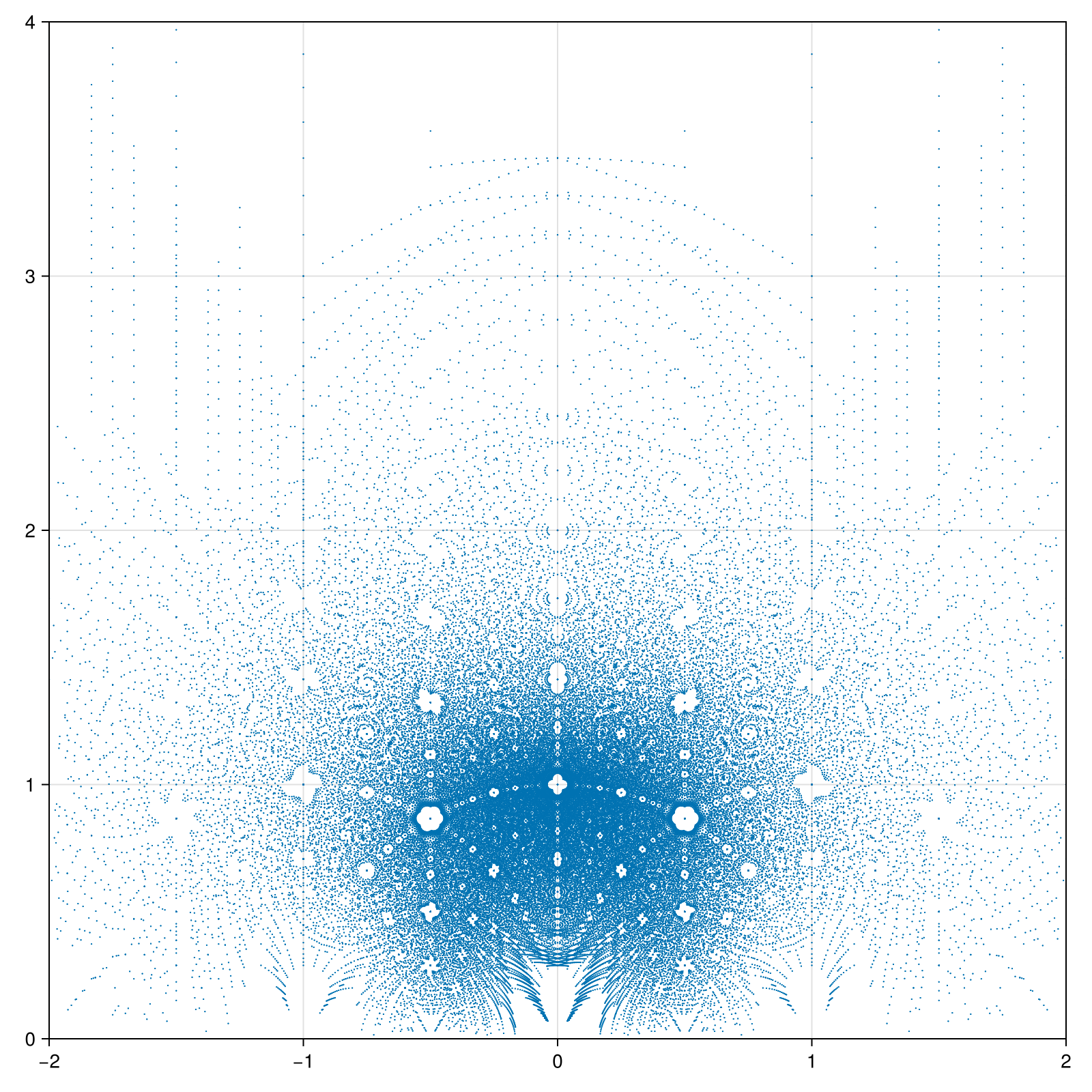}
\caption{Solutions of the SUSY conditions~(\ref{eq:orientifold_alpha}), (\ref{eq:orientifold_beta}) in the 
\(T^6/(\mathbb{Z}_2\times\mathbb{Z}_2')\) orientifold
model for \(-12 \leq e_0, e, m, m_0 \leq 12\).}
\label{fig:orientifold_N_12}
\end{figure}

\begin{figure}
\centering
\includegraphics[width=0.6\linewidth,height=\textheight,keepaspectratio]{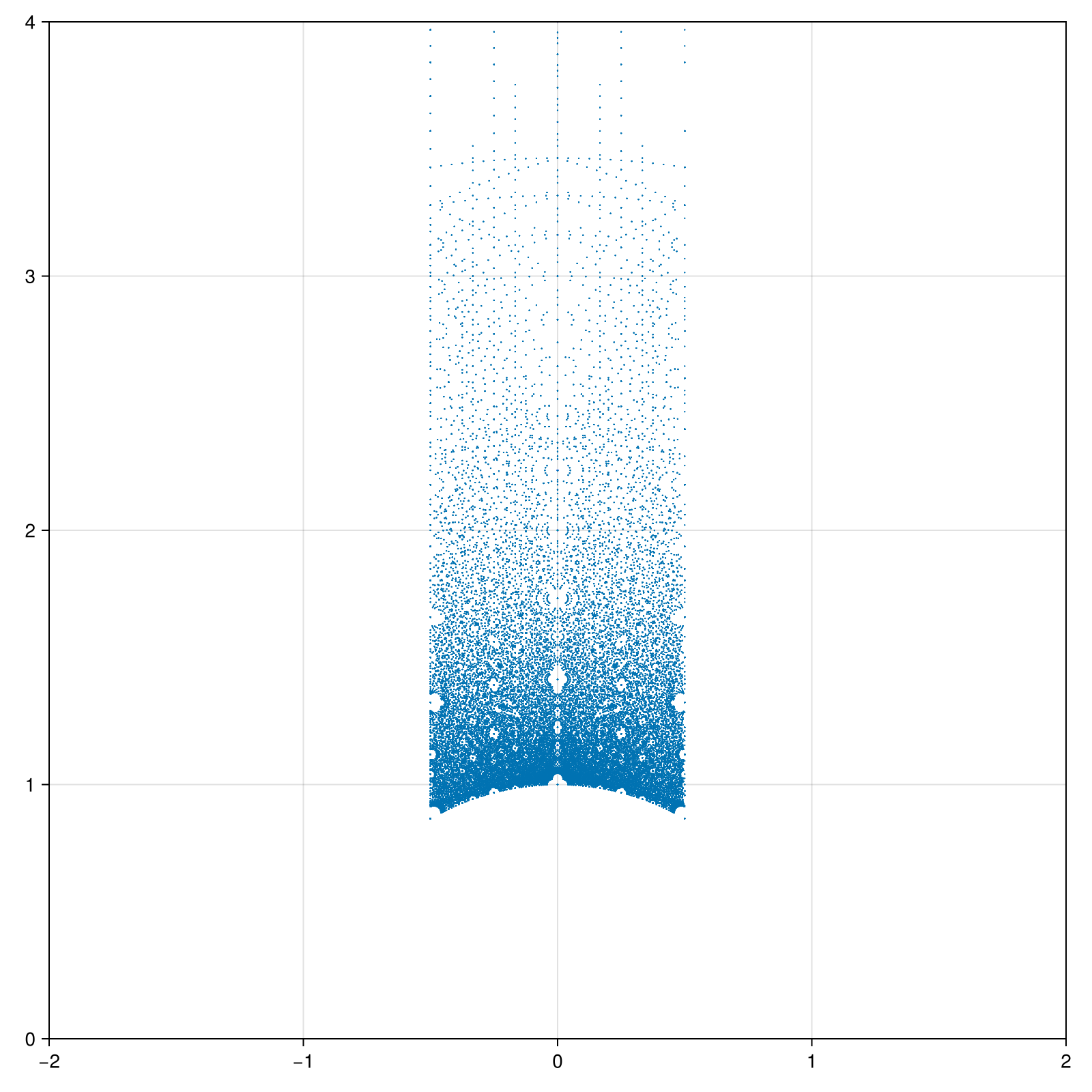}
\caption{Solutions of the SUSY conditions (\ref{eq:orientifold_alpha}), (\ref{eq:orientifold_beta}) in the 
fundamental domain.
}\label{fig:orientifold_N_12_fund}
\end{figure}

In what follows, we vary the fluxes in the region 
\(-N \leq e_0, e, m, m_0 \leq N\), and analyze the SUSY condition.
Figure~\ref{fig:orientifold_N_12}  shows the solutions of $u$ for $N=12$ in the upper half plane.
In Figure~\ref{fig:orientifold_N_12_fund}, we use the modular transformation such that  the points outside of the fundamental domain are transformed to the inside of the fundamental domain.
Figure~\ref{fig:orientifold_N_12}  as well as 
 Figure~\ref{fig:orientifold_N_12_fund} shows 
the straight line pattern $\alpha=$~constant in Eq.~(\ref{eq:orientifold_alpha}) and the circle pattern in Eqs.~(\ref{eq:orientifold_circ_m_m0}) and (\ref{eq:orientifold_circ_e_m}).

When we combine Eqs.~(\ref{eq:orientifold_alpha}) and (\ref{eq:orientifold_beta}), the solution $u$ is written by
\begin{equation}\label{eq:u-sol}
{u(k) = \frac{-B(k) - i\sqrt{-D(k)}}{2A(k)}}, \quad D(k) = B(k)^2 - 4A(k)C(k),
\end{equation}
where we denote a combination of fluxes by 
$k=(e_0,e,m,m_0)$, and 
\begin{equation}
 A(k)=em_0-m^2, \quad  B(k)=e_0m_0-em, \quad C(k)=e_0m-e^2.
\end{equation}
Thus, the solution belongs to the imaginary quadratic number field $\mathbb{Q}(\sqrt{-D(k)})$ \cite{DeWolfe:2004ns}.
Also, the solution $u$ satisfies the following holomorphic quadratic equation:
\begin{equation}\label{eq:moduli_quadratic_eq}
F(u; k) \equiv A(k)u^2 + B(k)u + C(k) = 0.
\end{equation} 
We require that its solution must satisfy 
${\rm Im}u>0$.
That means 
\begin{equation}
D(k) < 0.
\end{equation}
Furthermore, it is required 
\begin{equation}
A(k) < 0 {\qq{and} C(k) < 0}
\end{equation}
in order {for} Eqs.~(\ref{eq:orientifold_circ_m_m0}) and (\ref{eq:orientifold_circ_e_m}) to have solutions.

\subsection{Other types of analyses}
\label{sec:other}

In the previous subsection, we have derived the holomorphic quadratic equation (\ref{eq:moduli_quadratic_eq}) through the Gr\"obner method.
Here, we derive the same equation in different ways.

\subsubsection{Direct derivation}

From the SUSY condition (\ref{eq:SUSY-min}), we can derive the following equation:
\begin{equation}\label{eq:orientifold_cond_u_bar}
    \bar{u}({e + 2mu + m_0u^2}) = - {e_0 + 2eu + mu^2},
\end{equation}
and its conjugate equation,
\begin{equation}
    u ({e + 2m\bar{u} + m_0\bar{u}^2})= - {e_0 + 2e\bar{u} + m\bar{u}^2}.
\end{equation}
By combining these equations, we can obtain the following equation:
\begin{align}
    0 &= (e_0 + 3e u + 3m u^2 + m_0 u^3) \cdot [(em_0-m^2)u^2 + (e_0m_0-em)u + (e_0m-e^2)] \notag \\
    &= W(u)\cdot F(u).
\end{align}
If $W\neq 0$, we find the above quadratic equation, $F(u)=0$.
If $W(u)=0$, the SUSY condition can be written by
\begin{align}
    W_u=e + 2mu + m_0u^2=0.
\end{align}
That is a specific case of the quadratic equation.

\subsubsection{Introduction of additional field}

Here, we introduce a new field $t$ such that we write the superpotential by the following homogeneous one:
\begin{equation}\label{eq:orientifold_f}
\mathcal{W}(u,t) = m_0u^3 + 3mu^2t + 3eut^2 + e_0t^3, \quad W(u) = \mathcal{W}(u,1).
\end{equation} 
The new field $t$ may correspond to gauge degree of freedom to define the complex structure basis of $u$.

The SUSY condition (\ref{eq:SUSY-min}) is equivalent to
\begin{equation}
(u-\bar{u})\mathcal{W}_u + (t-\bar{t})\mathcal{W}_t - 3\mathcal{W} = 0.
\end{equation}
In addition, since $\mathcal{W}(u,t)$ is the homogeneous cubic function, we find the identity \(u\mathcal{W}_u + t\mathcal{W}_t = 3\mathcal{W}\).
By using it, the SUSY condition can be written by
\begin{equation}
\mathcal{W}_i \bar{u}^i=\mathcal{W}_u \bar{u}+ \mathcal{W}_t\bar{t} = 0,
\end{equation}
where we have denoted $u^i = (u, t)$.
Furthermore, by use of the identities $\mathcal{W}_{ij} u^j = 2\mathcal{W}_i$, $\mathcal{W}_{ijk} u^k = \mathcal{W}_{ij}$, the SUSY condition can be written
\begin{equation}\label{eq:Wijk_u_u_ubar}
    \mathcal{W}_{ijk} u^i u^j \bar{u}^k = 0, \qquad 
      \mathcal{W}_{ijk} u^i \bar{u}^j \bar{u}^k = 0.
\end{equation}
Note that $\mathcal{W}_{ijk}$ is real and symmetric for $i,j,k$.

Here, we define the following vector:
\begin{equation}
    w_i = \mathcal{W}_{ijk} u^j \bar{u}^k = \mathcal{W}_{ij} \bar{u}^j,
\end{equation}
and it satisfies 
\begin{equation}\label{eq:w_u_0_w_ubar_0}
    w_i u^i = 0, \quad w_i \bar{u}^i = 0,
\end{equation}
because of Eq.~(\ref{eq:Wijk_u_u_ubar}).
The vector $w_i$ is orthogonal to $u^i$ and $\bar{u}^i$, and $u^i$ and $\bar{u}^i$ can span the basis on $\mathbb{C}^2$.
That implies 
\begin{equation}
    w_i|_{t=1} = \mathcal{W}_{ij} (u, 1) \bar{u}^j = 0.
\end{equation}
Since $\bar{u}^i$ should have a non-trivial solution, it is required that
\begin{equation}
H(u,1) \equiv \vmqty{\mathcal{W}_{uu} & \mathcal{W}_{ut} \\ \mathcal{W}_{tu} & \mathcal{W}_{tt}}_{t=1}
= -36\bqty{(em_0-m^2)u^2 + (e_0m_0-em) u + (e_0m-e^2)} = 0.
\end{equation}
Note that 
\begin{equation}
F(u;k) = - \frac{H(u,1)}{36}.
\end{equation}
Thus, the SUSY condition ensures \(F(u;k)=0\) .

The condition $H(u,1)=0$ means that $(u,t)$ has a flat direction.
There is a degree of freedom to scale $u$ to match $t$.
When we fix the gauge degree of freedom $t$, $u$ can be fixed.
As this method shows, 
the meaning of the quadratic equation (\ref{eq:moduli_quadratic_eq}) is that there is a flat direction if we introduce the gauge degree of freedom $t$. 

\subsection{Modular symmetry}
\label{sec:modular}

$T^6/(\mathbb{Z}_2\times\mathbb{Z}_2')$ has the modular symmetry without flux background.
It is generated by $S$ and $T$ transformations, i.e.,
\begin{align}
    S:~u\to -\frac{1}{u}, \qquad 
    T:u\to u+1.
\end{align}
One can see these symmetric patterns in Figure~\ref{fig:orientifold_N_12}.
Indeed, the $S$ transformation corresponds to the replacement of fluxes,
\begin{align}
      e_0&\rightarrow-m_0, & e&\rightarrow m, & m&\rightarrow-e, & m_0&\rightarrow e_0, 
\end{align}
and $T$ transformation corresponds to the shift of fluxes,
\begin{align}
    e_0&\rightarrow e_0+3e+3m+m_0, & e&\rightarrow e+2m+m_0, & m&\rightarrow m+m_0, & m_0&\rightarrow m_0.
\end{align}

Here, we also comment that a straight line $\alpha=$ constant like Eq.~(\ref{eq:orientifold_alpha}) can be transformed to a circle like Eqs.~(\ref{eq:orientifold_circ_m_m0}), (\ref{eq:orientifold_circ_e_m}).
We parametrize the straight line by $u=\alpha(1+i\tan \theta)$.
We perform the $S$ transformation,
\begin{align}
S:~u \to u' & =-\frac{1}{u}  \notag \\ &=-\frac{1}{\alpha(1+i \tan \theta)}  \notag \\ 
&=\frac{-1}{2\alpha}\left( (\cos (2\theta) +1) -i \sin (2 \theta) \right).
\end{align}
$u'$ represents the circle centered at $-1/\alpha$ with the radius $1/|2\alpha|$.
This circle always passes through the origin, because 
the origin and $\beta \to \infty$ are $S$-dual to each other.
One can shift the center of circle by $u\to u+n$.
For example, when $1/(2\alpha)=$ integer, one can obtain a circle centered at the origin.
When $m=0$, the circle (\ref{eq:orientifold_circ_m_m0}) is written as 
\begin{equation}\label{eq:orientifold_circ_m=0}
 \alpha^2 + \beta^2 = -\frac{e}{m_0}.
\end{equation}
Similarly, the straight line $\beta=$ constant is transformed to the circle centered at $(\alpha,\beta)=(0,-1/\beta)$ with the radius $1/|2\beta|$.

\subsection{Distribution of modulus vacua}
\label{sec:vacua}

The resulting Figures \ref{fig:orientifold_N_12} and \ref{fig:orientifold_N_12_fund} show a characteristic pattern like the figure in Ref.~\cite{Denef:2004ze}.
Here, we examine these figures.

\subsubsection{Farey sequence}
\label{sec:Farey}

The solutions of Eqs.~(\ref{eq:orientifold_alpha}), (\ref{eq:orientifold_beta}) for \(-N \leq e_0, e, m, m_0 \leq N\) become specific sequence with $\alpha$ fixed.
Table \ref{tab:beta-orientifold-0} shows 
$\beta^2$ with $\alpha=0$ when we vary $N$ from 1 to 10.
The number in parentheses denotes the degeneracy of combinations $(e_0, e, m, m_0)$ leading to the same value of $\beta^2$.
We can find  values of $\beta^2$  between $\beta^2 \leq 1$ and $\beta^2 \geq 1$ in the table are transformed each other by the $S$ transformation, i.e., $\beta^2 \longleftrightarrow1/\beta^2$.
When we focus on $\beta^2 \leq 1$, these values correspond to the Farey sequence $F_N$, which is the set of irreducible rational numbers, $p,q$ with $0<p\leq q \leq 1$, where $p$ and $q$ are co-prime.
Suppose that $p_i/q_i$, $p_{i+1}/q_{i+1}$, and $p_{i+2}/q_{i+2}$ are three successive numbers in $F_N$.
They satisfy the following properties.
\begin{enumerate}
\item
 They satisfy $q_ip_{i+1}-q_{i+1}p_i=1$, and the following matrix:
  \begin{align}
      \begin{pmatrix}
          q_i & q_{i+1} \\
          p_i & p_{i+1}
      \end{pmatrix}
  \end{align}
belongs to $SL(2,\mathbb{Z})$.

\item
 These three successive numbers satisfy
 \begin{align}
     \frac{p_{i+1}}{q_{i+1}}=\frac{p_i+p_{i+2}}{q_i+q_{i+2}}.
 \end{align}
\end{enumerate}

\begin{longtable}{ll}
    \caption{$\beta^2$ sequence with $\alpha=0$ for \(-N \leq e_0, e, m, m_0 \leq N\). }
    \label{tab:beta-orientifold-0}\\ \hline
    $N$ & $\beta^2~(\alpha=0)$ \\ \hline
    \endfirsthead
    \hline
    $N$ & $\beta^2~(\alpha=0)$ \\ \hline
    \endhead
    \hline
    \endfoot
    \hline
    \endlastfoot
    \(1\) & \(\frac{1}{1}_{(8)}\) \\
 \(2\) & \(\frac{1}{2}_{(8)}\), \(\frac{1}{1}_{(24)}\),
 \(\frac{2}{1}_{(8)}\) \\
 \(3\) & \(\frac{1}{3}_{(8)}\), \(\frac{1}{2}_{(8)}\),
\(\frac{2}{3}_{(8)}\), \(\frac{1}{1}_{(48)}\), \(\frac{3}{2}_{(8)}\),
 \(\frac{2}{1}_{(8)}\), \(\frac{3}{1}_{(8)}\) \\
 \(4\) & \(\frac{1}{4}_{(8)}\), \(\frac{1}{3}_{(8)}\),
 \(\frac{1}{2}_{(24)}\), \(\frac{2}{3}_{(8)}\), \(\frac{3}{4}_{(8)}\),
 \(\frac{1}{1}_{(80)}\), \(\frac{4}{3}_{(8)}\), \(\frac{3}{2}_{(8)}\),
 \(\frac{2}{1}_{(24)}\), \(\frac{3}{1}_{(8)}\), \(\frac{4}{1}_{(8)}\) \\
 \(5\) & \(\frac{1}{5}_{(8)}\), \(\frac{1}{4}_{(8)}\),
 \(\frac{1}{3}_{(8)}\), \(\frac{2}{5}_{(8)}\), \(\frac{1}{2}_{(24)}\),
 \(\frac{3}{5}_{(8)}\), \(\frac{2}{3}_{(8)}\), \(\frac{3}{4}_{(8)}\),
 \(\frac{4}{5}_{(8)}\), \(\frac{1}{1}_{(120)}\), \(\frac{5}{4}_{(8)}\),
 \(\frac{4}{3}_{(8)}\), \(\frac{3}{2}_{(8)}\), \(\frac{5}{3}_{(8)}\), 
 \(\frac{2}{1}_{(24)}\), \(\frac{5}{2}_{(8)}\),\\ & \(\frac{3}{1}_{(8)}\), 
 \(\frac{4}{1}_{(8)}\), \(\frac{5}{1}_{(8)}\) \\
 \(6\) & \(\frac{1}{6}_{(8)}\), \(\frac{1}{5}_{(8)}\),
 \(\frac{1}{4}_{(8)}\), \(\frac{1}{3}_{(24)}\), \(\frac{2}{5}_{(8)}\),
 \(\frac{1}{2}_{(48)}\), \(\frac{3}{5}_{(8)}\), \(\frac{2}{3}_{(24)}\),
 \(\frac{3}{4}_{(8)}\), \(\frac{4}{5}_{(8)}\), \(\frac{5}{6}_{(8)}\),
 \(\frac{1}{1}_{(168)}\), \(\frac{6}{5}_{(8)}\), \(\frac{5}{4}_{(8)}\),
 \(\frac{4}{3}_{(8)}\), \(\frac{3}{2}_{(24)}\), \\ & \(\frac{5}{3}_{(8)}\), 
 \(\frac{2}{1}_{(48)}\), \(\frac{5}{2}_{(8)}\), \(\frac{3}{1}_{(24)}\),
 \(\frac{4}{1}_{(8)}\), \(\frac{5}{1}_{(8)}\), \(\frac{6}{1}_{(8)}\) \\
 \(7\) & \(\frac{1}{7}_{(8)}\), \(\frac{1}{6}_{(8)}\),
\(\frac{1}{5}_{(8)}\), \(\frac{1}{4}_{(8)}\), \(\frac{2}{7}_{(8)}\),
 \(\frac{1}{3}_{(24)}\), \(\frac{2}{5}_{(8)}\), \(\frac{3}{7}_{(8)}\),
 \(\frac{1}{2}_{(48)}\), \(\frac{4}{7}_{(8)}\), \(\frac{3}{5}_{(8)}\),
 \(\frac{2}{3}_{(24)}\), \(\frac{5}{7}_{(8)}\), \(\frac{3}{4}_{(8)}\),
 \(\frac{4}{5}_{(8)}\), \(\frac{5}{6}_{(8)}\), \\ &\(\frac{6}{7}_{(8)}\),
 \(\frac{1}{1}_{(224)}\), \(\frac{7}{6}_{(8)}\), \(\frac{6}{5}_{(8)}\),
 \(\frac{5}{4}_{(8)}\), \(\frac{4}{3}_{(8)}\), \(\frac{7}{5}_{(8)}\),
 \(\frac{3}{2}_{(24)}\), \(\frac{5}{3}_{(8)}\), \(\frac{7}{4}_{(8)}\),
 \(\frac{2}{1}_{(48)}\), \(\frac{7}{3}_{(8)}\), \(\frac{5}{2}_{(8)}\),
 \(\frac{3}{1}_{(24)}\), \(\frac{7}{2}_{(8)}\), \(\frac{4}{1}_{(8)}\), \\ &
 \(\frac{5}{1}_{(8)}\), \(\frac{6}{1}_{(8)}\), \(\frac{7}{1}_{(8)}\) \\
 \(8\) & \(\frac{1}{8}_{(8)}\), \(\frac{1}{7}_{(8)}\),
 \(\frac{1}{6}_{(8)}\), \(\frac{1}{5}_{(8)}\), \(\frac{1}{4}_{(24)}\),
 \(\frac{2}{7}_{(8)}\), \(\frac{1}{3}_{(24)}\), \(\frac{3}{8}_{(8)}\),
 \(\frac{2}{5}_{(8)}\), \(\frac{3}{7}_{(8)}\), \(\frac{1}{2}_{(80)}\),
 \(\frac{4}{7}_{(8)}\), \(\frac{3}{5}_{(8)}\), \(\frac{5}{8}_{(8)}\),
 \(\frac{2}{3}_{(24)}\), \(\frac{5}{7}_{(8)}\), \\ &\(\frac{3}{4}_{(24)}\),
 \(\frac{4}{5}_{(8)}\), \(\frac{5}{6}_{(8)}\), \(\frac{6}{7}_{(8)}\),
 \(\frac{7}{8}_{(8)}\), \(\frac{1}{1}_{(288)}\), \(\frac{8}{7}_{(8)}\),
 \(\frac{7}{6}_{(8)}\), \(\frac{6}{5}_{(8)}\), \(\frac{5}{4}_{(8)}\),
 \(\frac{4}{3}_{(24)}\), \(\frac{7}{5}_{(8)}\), \(\frac{3}{2}_{(24)}\),
 \(\frac{8}{5}_{(8)}\), \(\frac{5}{3}_{(8)}\), \(\frac{7}{4}_{(8)}\), \\ &
 \(\frac{2}{1}_{(80)}\), \(\frac{7}{3}_{(8)}\), \(\frac{5}{2}_{(8)}\),
 \(\frac{8}{3}_{(8)}\), \(\frac{3}{1}_{(24)}\), \(\frac{7}{2}_{(8)}\),
 \(\frac{4}{1}_{(24)}\), \(\frac{5}{1}_{(8)}\), \(\frac{6}{1}_{(8)}\),
 \(\frac{7}{1}_{(8)}\), \(\frac{8}{1}_{(8)}\) \\
 \(9\) & \(\frac{1}{9}_{(8)}\), \(\frac{1}{8}_{(8)}\),
 \(\frac{1}{7}_{(8)}\), \(\frac{1}{6}_{(8)}\), \(\frac{1}{5}_{(8)}\),
 \(\frac{2}{9}_{(8)}\), \(\frac{1}{4}_{(24)}\), \(\frac{2}{7}_{(8)}\),
 \(\frac{1}{3}_{(48)}\), \(\frac{3}{8}_{(8)}\), \(\frac{2}{5}_{(8)}\),
 \(\frac{3}{7}_{(8)}\), \(\frac{4}{9}_{(8)}\), \(\frac{1}{2}_{(80)}\),
 \(\frac{5}{9}_{(8)}\), \(\frac{4}{7}_{(8)}\), \\ & \(\frac{3}{5}_{(8)}\),
 \(\frac{5}{8}_{(8)}\), \(\frac{2}{3}_{(48)}\), \(\frac{5}{7}_{(8)}\),
 \(\frac{3}{4}_{(24)}\), \(\frac{7}{9}_{(8)}\), \(\frac{4}{5}_{(8)}\),
 \(\frac{5}{6}_{(8)}\), \(\frac{6}{7}_{(8)}\), \(\frac{7}{8}_{(8)}\),
 \(\frac{8}{9}_{(8)}\), \(\frac{1}{1}_{(360)}\), \(\frac{9}{8}_{(8)}\),
 \(\frac{8}{7}_{(8)}\), \(\frac{7}{6}_{(8)}\), \(\frac{6}{5}_{(8)}\), \\ &
 \(\frac{5}{4}_{(8)}\), \(\frac{9}{7}_{(8)}\), \(\frac{4}{3}_{(24)}\),
 \(\frac{7}{5}_{(8)}\), \(\frac{3}{2}_{(48)}\), \(\frac{8}{5}_{(8)}\),
 \(\frac{5}{3}_{(8)}\), \(\frac{7}{4}_{(8)}\), \(\frac{9}{5}_{(8)}\),
 \(\frac{2}{1}_{(80)}\), \(\frac{9}{4}_{(8)}\), \(\frac{7}{3}_{(8)}\),
 \(\frac{5}{2}_{(8)}\), \(\frac{8}{3}_{(8)}\), \(\frac{3}{1}_{(48)}\),
 \(\frac{7}{2}_{(8)}\), \\ & \(\frac{4}{1}_{(24)}\), \(\frac{9}{2}_{(8)}\),
 \(\frac{5}{1}_{(8)}\), \(\frac{6}{1}_{(8)}\), \(\frac{7}{1}_{(8)}\),
 \(\frac{8}{1}_{(8)}\), \(\frac{9}{1}_{(8)}\) \\
 \(10\) & \(\frac{1}{10}_{(8)}\), \(\frac{1}{9}_{(8)}\),
 \(\frac{1}{8}_{(8)}\), \(\frac{1}{7}_{(8)}\), \(\frac{1}{6}_{(8)}\),
 \(\frac{1}{5}_{(24)}\), \(\frac{2}{9}_{(8)}\), \(\frac{1}{4}_{(24)}\),
 \(\frac{2}{7}_{(8)}\), \(\frac{3}{10}_{(8)}\), \(\frac{1}{3}_{(48)}\),
 \(\frac{3}{8}_{(8)}\), \(\frac{2}{5}_{(24)}\), \(\frac{3}{7}_{(8)}\),
 \(\frac{4}{9}_{(8)}\), \\ &\(\frac{1}{2}_{(120)}\),  \(\frac{5}{9}_{(8)}\),
 \(\frac{4}{7}_{(8)}\), \(\frac{3}{5}_{(24)}\), \(\frac{5}{8}_{(8)}\),
 \(\frac{2}{3}_{(48)}\), \(\frac{7}{10}_{(8)}\), \(\frac{5}{7}_{(8)}\),
 \(\frac{3}{4}_{(24)}\), \(\frac{7}{9}_{(8)}\), \(\frac{4}{5}_{(24)}\),
 \(\frac{5}{6}_{(8)}\), \(\frac{6}{7}_{(8)}\), \(\frac{7}{8}_{(8)}\),
 \(\frac{8}{9}_{(8)}\), \\ & \(\frac{9}{10}_{(8)}\), \(\frac{1}{1}_{(440)}\),
 \(\frac{10}{9}_{(8)}\), \(\frac{9}{8}_{(8)}\), \(\frac{8}{7}_{(8)}\),
 \(\frac{7}{6}_{(8)}\), \(\frac{6}{5}_{(8)}\), \(\frac{5}{4}_{(24)}\),
 \(\frac{9}{7}_{(8)}\), \(\frac{4}{3}_{(24)}\), \(\frac{7}{5}_{(8)}\),
 \(\frac{10}{7}_{(8)}\), \(\frac{3}{2}_{(48)}\), \(\frac{8}{5}_{(8)}\),
 \(\frac{5}{3}_{(24)}\), \\ & \(\frac{7}{4}_{(8)}\), \(\frac{9}{5}_{(8)}\),
 \(\frac{2}{1}_{(120)}\), \(\frac{9}{4}_{(8)}\), \(\frac{7}{3}_{(8)}\),
 \(\frac{5}{2}_{(24)}\), \(\frac{8}{3}_{(8)}\), \(\frac{3}{1}_{(48)}\),
 \(\frac{10}{3}_{(8)}\), \(\frac{7}{2}_{(8)}\), \(\frac{4}{1}_{(24)}\),
 \(\frac{9}{2}_{(8)}\), \(\frac{5}{1}_{(24)}\), \(\frac{6}{1}_{(8)}\),
 \(\frac{7}{1}_{(8)}\), \\ & \(\frac{8}{1}_{(8)}\), \(\frac{9}{1}_{(8)}\),
 \(\frac{10}{1}_{(8)}\) \\
\end{longtable}

The reason why the Farey sequence appears is as follows.
We define the following matrix:
\begin{align}
    M(k)=
    \begin{pmatrix}
        e_0 & m \\
        e &m_0
    \end{pmatrix}.
\end{align}
Its determinant is equal to $B(k)=\det M(k)$.
When ${\rm Re}~u=0$, we find $B(k)=\det M(k)=0$.
Hence, we can write
\begin{align}
    \begin{pmatrix}
        e_0 \\ e
    \end{pmatrix}
    =\lambda\begin{pmatrix}
        r \\ s
    \end{pmatrix}, \qquad
     \begin{pmatrix}
        m \\ m_0
    \end{pmatrix}
    =\mu\begin{pmatrix}
        r \\ s
    \end{pmatrix}, 
\end{align}
where $\lambda,\mu,r,s$ are integers and 
$r$ and $s$ are coprime.
By use of them, we can write
\begin{align}
    A(k) = \mu (\lambda s^2 - \mu r^2), \qquad 
    C(k)  = -\lambda (\lambda s^2 - \mu r^2).
\end{align}
Then, we obtain 
\begin{equation}
\label{eq:lambda/mu}
    ({\rm Im} ~u)^2 =\beta^2= \frac{C(k)}{A(k)} = - \frac{\lambda}{\mu},
\end{equation}
which is irreducible rational number, because $\lambda$ and $\mu$ are coprime.
Obviously, $|\mu|$ satisfies
\begin{align}
    |\mu| \leq {\rm max}(|m|,|m_0|)\leq N.
\end{align}
Thus, the denominator of the irreducible rational number in Eq.~(\ref{eq:lambda/mu}) is less than or equal to $N$.
On the other hand, any irreducible rational number $\beta^2=p/q$ with $q\lneq N$ can be realized by use of flux combinations $(e_0,e,m,m_0)$.
For example, let us set 
$(e_0,e,m,m_0)=(-p,0,q,0)$, i.e.,
\begin{align}
    M(k)=
    \begin{pmatrix}
        -p & q \\
        0 & 0
    \end{pmatrix}.
\end{align}
Then, we can realize $\beta^2=p/q$. 
{This is the reason why the values of} $\beta^2$ in Table~\ref{tab:beta-orientifold-0} correspond to the Farey sequence.

The degeneracies of rational numbers, which can appear in $F_N$ with smaller $N$ are larger, because $F_N$ sequence always include $F_{N-1}$.
For example, the degeneracy is highest in 
$\beta^2=1$, followed by $\beta^2=1/2$ and $2$, and then
$\beta^2=1/3$, 2/3, 3/2 and $3$.
The degeneracies of rational numbers, which appear in $F_N$ for the first time, but do not appear in $F_{N-1}$, 
are the same.

For $\beta^2=1$, we need $\lambda=-\mu$, that is, $e_0=-m$ and $e=-m_0$.
Both $e_0$ and $e$ vary $-N \leq e_0, e \leq N$ except 
$e_0=e=m_0=m=0$.
Therefore, the degeneracy of $u=i$ is equal to $(2N+1)^2-1$.
Similarly, for $\beta^2=1/2$ (2), we need $2\lambda=-\mu$ ($\lambda=-2\mu$), that is $2e_0=-m$ and $2e=-m_0$ ($e_0=-2m$ and $e=-2m_0$).
Both $e_0$ and $e$ vary $-N/2 \leq e_0, e \leq N/2$ except $e_0=e=m_0=m=0$.
Hence, the degeneracy of  $\beta^2=1/2$ and 2
is equal to $(2\lfloor N/2 \rfloor +1)^2-1$, 
where $\lfloor x \rfloor$ denotes the floor function, i.e., the integer part of $x$.
Similarly, the degeneracy of $\beta^2=1/3$, 2/3, 3/2 and $3$ is given by $(2\lfloor N/3 \rfloor +1)^2-1$. 
As a result, the degeneracies $d$ are written by
\begin{align}
    d=(2m+1)^2-1=8T_m=8,24,48,80,120,168,224,288,360,440,\cdots,
\end{align}
where $T_m=m(m+1)/2$ denotes the triangular number.

This result on the degeneracy is important in modular flavor models 
as emphasized in Ref.~\cite{Ishiguro:2020tmo}.
In particular, a small deviation around the fixed points $u=i$ and $e^{\pi i/3}$ and large ${\rm Im}~u$ are interesting to realize hierarchical fermion masses \cite{Feruglio:2021dte,Novichkov:2021evw,Petcov:2022fjf,Kikuchi:2023cap,Abe:2023ilq,Kikuchi:2023jap,Abe:2023qmr,Petcov:2023vws,Abe:2023dvr,deMedeirosVarzielas:2023crv,Kikuchi:2023fpl}.
However, there are voids around $u=i$ and $e^{\pi i /3}$.
Flux stabilization of $u$ cannot realize such a small deviation from the fixed points $u=i$ and $e^{\pi i /3}$.
We need other effects to realize such deviation \cite{Ishiguro:2022pde,Kobayashi:2023spx}, although large ${\rm Im}~ u$ can be realized.
For example, we can realize $q=e^{-2\pi {\rm Im}~u}=0.04$ when ${\rm Im}~u=2$.
Hence, large ${\rm Im}~u$ seems to be realized more readily than small deviations from $u=i$ and $e^{2\pi i/3 }$.

Similarly, Table \ref{tab:beta-orientifold-1/2} shows $\beta^2$ at $\alpha=1/2$.
Results show a specific sequence whose properties are similar to the above results in Table~\ref{tab:beta-orientifold-0}.
For fixed $N$, the variety of $\beta^2$ with $\alpha=1/2$ is greater than those with $\alpha=0$.
The degeneracy is highest in $\beta^2=3/4$, followed by $\beta^2=7/4$.
Note that $\beta^2=3/4$ corresponds to one of the fixed points of modular symmetry, $u=e^{\pi i/3}$.

\begin{longtable}{ll}
    \caption{$\beta^2$ sequence with $\alpha=1/2$ for \(-N \leq e_0, e, m, m_0 \leq N\). }
    \label{tab:beta-orientifold-1/2}\\ \hline
    $N$ & $\beta^2~(\alpha=\frac{1}{2})$ \\ \hline
    \endfirsthead
    \hline
    $N$ & $\beta^2~(\alpha=\frac{1}{2})$ \\ \hline
    \endhead
    \hline
    \endfoot
    \hline
    \endlastfoot
\(1\) & \(\frac{3}{4}_{(6)}\), \(\frac{7}{4}_{(2)}\) \\
\(2\) & \(\frac{1}{4}_{(6)}\), \(\frac{3}{4}_{(18)}\),
\(\frac{5}{4}_{(2)}\), \(\frac{7}{4}_{(6)}\), \(\frac{11}{4}_{(2)}\) \\
\(3\) & \(\frac{1}{12}_{(6)}\), \(\frac{1}{4}_{(8)}\),
\(\frac{5}{12}_{(6)}\), \(\frac{3}{4}_{(36)}\), \(\frac{13}{12}_{(2)}\),
\(\frac{5}{4}_{(4)}\), \(\frac{17}{12}_{(2)}\), \(\frac{7}{4}_{(12)}\),
\(\frac{9}{4}_{(2)}\), \(\frac{11}{4}_{(4)}\), \(\frac{15}{4}_{(2)}\) \\
\(4\) & \(\frac{1}{12}_{(8)}\), \(\frac{1}{4}_{(20)}\),
\(\frac{5}{12}_{(6)}\), \(\frac{1}{2}_{(6)}\), \(\frac{3}{4}_{(60)}\),
\(\frac{1}{1}_{(2)}\), \(\frac{13}{12}_{(4)}\), \(\frac{5}{4}_{(8)}\),
\(\frac{17}{12}_{(2)}\), \(\frac{3}{2}_{(2)}\), \(\frac{7}{4}_{(20)}\),
\(\frac{25}{12}_{(2)}\), \(\frac{9}{4}_{(2)}\), \(\frac{11}{4}_{(8)}\), \\ &
\(\frac{15}{4}_{(4)}\), \(\frac{19}{4}_{(2)}\) \\
\(5\) & \(\frac{1}{12}_{(8)}\), \(\frac{3}{20}_{(6)}\),
\(\frac{1}{4}_{(22)}\), \(\frac{7}{20}_{(6)}\), \(\frac{5}{12}_{(8)}\),
\(\frac{1}{2}_{(6)}\), \(\frac{11}{20}_{(6)}\), \(\frac{3}{4}_{(90)}\),
\(\frac{19}{20}_{(2)}\), \(\frac{1}{1}_{(4)}\), \(\frac{13}{12}_{(6)}\),
\(\frac{23}{20}_{(2)}\), \(\frac{5}{4}_{(12)}\), \\ &
\(\frac{27}{20}_{(2)}\), \(\frac{17}{12}_{(4)}\), \(\frac{3}{2}_{(2)}\),
\(\frac{31}{20}_{(2)}\), \(\frac{7}{4}_{(30)}\), \(\frac{2}{1}_{(2)}\),
\(\frac{25}{12}_{(2)}\), \(\frac{9}{4}_{(4)}\), \(\frac{29}{12}_{(2)}\),
\(\frac{11}{4}_{(12)}\), \(\frac{13}{4}_{(2)}\), \(\frac{15}{4}_{(6)}\),
\(\frac{19}{4}_{(4)}\), \(\frac{23}{4}_{(2)}\) \\
\(6\) & \(\frac{1}{12}_{(20)}\), \(\frac{3}{20}_{(6)}\),
\(\frac{1}{4}_{(40)}\), \(\frac{7}{20}_{(6)}\), \(\frac{5}{12}_{(18)}\),
\(\frac{1}{2}_{(6)}\), \(\frac{11}{20}_{(6)}\), \(\frac{7}{12}_{(6)}\),
\(\frac{3}{4}_{(126)}\), \(\frac{11}{12}_{(2)}\),
\(\frac{19}{20}_{(4)}\), \(\frac{1}{1}_{(6)}\),
\(\frac{13}{12}_{(10)}\), \\ & \(\frac{23}{20}_{(2)}\),
\(\frac{5}{4}_{(18)}\), \(\frac{27}{20}_{(2)}\),
\(\frac{17}{12}_{(6)}\), \(\frac{3}{2}_{(2)}\), \(\frac{31}{20}_{(2)}\),
\(\frac{19}{12}_{(2)}\), \(\frac{7}{4}_{(42)}\),
\(\frac{39}{20}_{(2)}\), \(\frac{2}{1}_{(2)}\), \(\frac{25}{12}_{(2)}\),
\(\frac{9}{4}_{(6)}\), \(\frac{29}{12}_{(2)}\), \(\frac{11}{4}_{(18)}\),  \\ &
\(\frac{13}{4}_{(2)}\), \(\frac{15}{4}_{(10)}\), \(\frac{19}{4}_{(6)}\),
\(\frac{23}{4}_{(4)}\), \(\frac{27}{4}_{(2)}\) \\
\(7\) & \(\frac{1}{28}_{(6)}\), \(\frac{1}{12}_{(22)}\),
\(\frac{3}{20}_{(8)}\), \(\frac{5}{28}_{(6)}\), \(\frac{1}{4}_{(44)}\),
\(\frac{9}{28}_{(6)}\), \(\frac{7}{20}_{(6)}\), \(\frac{5}{12}_{(20)}\),
\(\frac{13}{28}_{(6)}\), \(\frac{1}{2}_{(8)}\), \(\frac{11}{20}_{(6)}\),
\(\frac{7}{12}_{(6)}\), \(\frac{17}{28}_{(6)}\),
\\ & \(\frac{3}{4}_{(168)}\), \(\frac{25}{28}_{(2)}\),
\(\frac{11}{12}_{(4)}\), \(\frac{19}{20}_{(6)}\), \(\frac{1}{1}_{(6)}\),
\(\frac{29}{28}_{(2)}\), \(\frac{13}{12}_{(12)}\),
\(\frac{23}{20}_{(4)}\), \(\frac{33}{28}_{(2)}\),
\(\frac{5}{4}_{(22)}\), \(\frac{37}{28}_{(2)}\),
\(\frac{27}{20}_{(2)}\), \(\frac{17}{12}_{(8)}\),  \\ & 
\(\frac{41}{28}_{(2)}\), \(\frac{3}{2}_{(4)}\), \(\frac{31}{20}_{(2)}\),
\(\frac{19}{12}_{(2)}\), \(\frac{45}{28}_{(2)}\),
\(\frac{7}{4}_{(56)}\), \(\frac{23}{12}_{(2)}\),
\(\frac{39}{20}_{(2)}\), \(\frac{2}{1}_{(2)}\), \(\frac{25}{12}_{(4)}\),
\(\frac{43}{20}_{(2)}\), \(\frac{9}{4}_{(8)}\), \(\frac{29}{12}_{(2)}\),
\(\frac{5}{2}_{(2)}\), \\ &  \(\frac{11}{4}_{(24)}\), \(\frac{37}{12}_{(2)}\),
\(\frac{13}{4}_{(4)}\), \(\frac{15}{4}_{(14)}\), \(\frac{17}{4}_{(2)}\),
\(\frac{19}{4}_{(8)}\), \(\frac{23}{4}_{(6)}\), \(\frac{27}{4}_{(4)}\),
\(\frac{31}{4}_{(2)}\) \\
\(8\) & \(\frac{1}{28}_{(6)}\), \(\frac{1}{12}_{(24)}\),
\(\frac{1}{8}_{(6)}\), \(\frac{3}{20}_{(8)}\), \(\frac{5}{28}_{(6)}\),
\(\frac{1}{4}_{(68)}\), \(\frac{9}{28}_{(6)}\), \(\frac{7}{20}_{(8)}\),
\(\frac{3}{8}_{(6)}\), \(\frac{5}{12}_{(22)}\), \(\frac{13}{28}_{(6)}\),
\(\frac{1}{2}_{(18)}\), \(\frac{11}{20}_{(6)}\),  \\ & \(\frac{7}{12}_{(6)}\),
\(\frac{17}{28}_{(6)}\), \(\frac{5}{8}_{(6)}\), \(\frac{3}{4}_{(216)}\),
\(\frac{7}{8}_{(2)}\), \(\frac{25}{28}_{(4)}\), \(\frac{11}{12}_{(6)}\),
\(\frac{19}{20}_{(6)}\), \(\frac{1}{1}_{(10)}\),
\(\frac{29}{28}_{(2)}\), \(\frac{13}{12}_{(14)}\),
\(\frac{9}{8}_{(2)}\), \(\frac{23}{20}_{(4)}\), \(\frac{33}{28}_{(2)}\),  \\ &
\(\frac{5}{4}_{(30)}\), \(\frac{37}{28}_{(2)}\),
\(\frac{27}{20}_{(4)}\), \(\frac{11}{8}_{(2)}\),
\(\frac{17}{12}_{(10)}\), \(\frac{41}{28}_{(2)}\),
\(\frac{3}{2}_{(6)}\), \(\frac{31}{20}_{(2)}\), \(\frac{19}{12}_{(2)}\),
\(\frac{45}{28}_{(2)}\), \(\frac{13}{8}_{(2)}\), \(\frac{7}{4}_{(72)}\),
\(\frac{53}{28}_{(2)}\), \\ &  \(\frac{23}{12}_{(2)}\),
\(\frac{39}{20}_{(2)}\), \(\frac{2}{1}_{(2)}\), \(\frac{25}{12}_{(6)}\),
\(\frac{43}{20}_{(2)}\), \(\frac{9}{4}_{(10)}\),
\(\frac{47}{20}_{(2)}\), \(\frac{29}{12}_{(4)}\), \(\frac{5}{2}_{(2)}\),
\(\frac{11}{4}_{(32)}\), \(\frac{37}{12}_{(2)}\),
\(\frac{13}{4}_{(4)}\), \(\frac{41}{12}_{(2)}\),
\\ & \(\frac{15}{4}_{(18)}\), \(\frac{17}{4}_{(2)}\),
\(\frac{19}{4}_{(12)}\), \(\frac{23}{4}_{(8)}\), \(\frac{27}{4}_{(6)}\),
\(\frac{31}{4}_{(4)}\), \(\frac{35}{4}_{(2)}\) \\
\(9\) & \(\frac{1}{28}_{(8)}\), \(\frac{1}{12}_{(42)}\),
\(\frac{1}{8}_{(6)}\), \(\frac{3}{20}_{(8)}\), \(\frac{5}{28}_{(6)}\),
\(\frac{7}{36}_{(6)}\), \(\frac{1}{4}_{(72)}\), \(\frac{11}{36}_{(6)}\),
\(\frac{9}{28}_{(6)}\), \(\frac{7}{20}_{(8)}\), \(\frac{3}{8}_{(6)}\),
\(\frac{5}{12}_{(38)}\), \(\frac{13}{28}_{(6)}\), \\ &
\(\frac{1}{2}_{(18)}\), \(\frac{19}{36}_{(6)}\),
\(\frac{11}{20}_{(8)}\), \(\frac{7}{12}_{(6)}\),
\(\frac{17}{28}_{(6)}\), \(\frac{5}{8}_{(6)}\), \(\frac{23}{36}_{(6)}\),
\(\frac{3}{4}_{(270)}\), \(\frac{31}{36}_{(2)}\), \(\frac{7}{8}_{(4)}\),
\(\frac{25}{28}_{(6)}\), \(\frac{11}{12}_{(6)}\),
\(\frac{19}{20}_{(6)}\), \\ & \(\frac{35}{36}_{(2)}\),
\(\frac{1}{1}_{(12)}\), \(\frac{29}{28}_{(4)}\),
\(\frac{13}{12}_{(20)}\), \(\frac{9}{8}_{(2)}\),
\(\frac{23}{20}_{(6)}\), \(\frac{33}{28}_{(2)}\),
\(\frac{43}{36}_{(2)}\), \(\frac{5}{4}_{(36)}\),
\(\frac{47}{36}_{(2)}\), \(\frac{37}{28}_{(2)}\),
\(\frac{27}{20}_{(4)}\), \(\frac{11}{8}_{(2)}\),
\\ & \(\frac{17}{12}_{(14)}\), \(\frac{41}{28}_{(2)}\),
\(\frac{3}{2}_{(6)}\), \(\frac{55}{36}_{(2)}\), \(\frac{31}{20}_{(4)}\),
\(\frac{19}{12}_{(2)}\), \(\frac{45}{28}_{(2)}\),
\(\frac{13}{8}_{(2)}\), \(\frac{59}{36}_{(2)}\), \(\frac{7}{4}_{(90)}\),
\(\frac{15}{8}_{(2)}\), \(\frac{53}{28}_{(2)}\),
\(\frac{23}{12}_{(2)}\), \\ & \(\frac{39}{20}_{(2)}\), \(\frac{2}{1}_{(4)}\),
\(\frac{57}{28}_{(2)}\), \(\frac{25}{12}_{(6)}\),
\(\frac{43}{20}_{(2)}\), \(\frac{9}{4}_{(14)}\),
\(\frac{47}{20}_{(2)}\), \(\frac{29}{12}_{(4)}\), \(\frac{5}{2}_{(2)}\),
\(\frac{51}{20}_{(2)}\), \(\frac{11}{4}_{(40)}\), \(\frac{3}{1}_{(2)}\),
\(\frac{37}{12}_{(2)}\), \(\frac{13}{4}_{(6)}\),
\\ & \(\frac{41}{12}_{(2)}\), \(\frac{15}{4}_{(22)}\),
\(\frac{17}{4}_{(4)}\), \(\frac{19}{4}_{(14)}\), \(\frac{21}{4}_{(2)}\),
\(\frac{23}{4}_{(10)}\), \(\frac{27}{4}_{(8)}\), \(\frac{31}{4}_{(6)}\),
\(\frac{35}{4}_{(4)}\), \(\frac{39}{4}_{(2)}\) \\
\(10\) & \(\frac{1}{28}_{(8)}\), \(\frac{1}{20}_{(6)}\),
\(\frac{1}{12}_{(44)}\), \(\frac{1}{8}_{(6)}\), \(\frac{3}{20}_{(20)}\),
\(\frac{5}{28}_{(8)}\), \(\frac{7}{36}_{(6)}\), \(\frac{1}{4}_{(102)}\),
\(\frac{11}{36}_{(6)}\), \(\frac{9}{28}_{(6)}\),
\(\frac{7}{20}_{(18)}\), \(\frac{3}{8}_{(6)}\), \(\frac{5}{12}_{(40)}\),  \\ & 
\(\frac{9}{20}_{(6)}\), \(\frac{13}{28}_{(6)}\), \(\frac{1}{2}_{(20)}\),
\(\frac{19}{36}_{(6)}\), \(\frac{11}{20}_{(18)}\),
\(\frac{7}{12}_{(6)}\), \(\frac{17}{28}_{(6)}\), \(\frac{5}{8}_{(6)}\),
\(\frac{23}{36}_{(6)}\), \(\frac{13}{20}_{(6)}\),
\(\frac{3}{4}_{(330)}\), \(\frac{17}{20}_{(2)}\),
\(\frac{31}{36}_{(4)}\), \\ & \(\frac{7}{8}_{(6)}\), \(\frac{25}{28}_{(6)}\),
\(\frac{11}{12}_{(6)}\), \(\frac{19}{20}_{(10)}\),
\(\frac{35}{36}_{(2)}\), \(\frac{1}{1}_{(14)}\),
\(\frac{29}{28}_{(4)}\), \(\frac{21}{20}_{(2)}\),
\(\frac{13}{12}_{(24)}\), \(\frac{9}{8}_{(2)}\),
\(\frac{23}{20}_{(8)}\), \(\frac{33}{28}_{(4)}\),
\(\frac{43}{36}_{(2)}\), \\ &  \(\frac{5}{4}_{(46)}\),
\(\frac{47}{36}_{(2)}\), \(\frac{37}{28}_{(2)}\),
\(\frac{27}{20}_{(6)}\), \(\frac{11}{8}_{(2)}\),
\(\frac{17}{12}_{(16)}\), \(\frac{29}{20}_{(2)}\),
\(\frac{41}{28}_{(2)}\), \(\frac{3}{2}_{(8)}\), \(\frac{55}{36}_{(2)}\),
\(\frac{31}{20}_{(6)}\), \(\frac{19}{12}_{(2)}\),
\(\frac{45}{28}_{(2)}\), \\ &  \(\frac{13}{8}_{(2)}\),
\(\frac{59}{36}_{(2)}\), \(\frac{33}{20}_{(2)}\),
\(\frac{7}{4}_{(110)}\), \(\frac{67}{36}_{(2)}\),
\(\frac{15}{8}_{(2)}\), \(\frac{53}{28}_{(2)}\),
\(\frac{23}{12}_{(2)}\), \(\frac{39}{20}_{(2)}\), \(\frac{2}{1}_{(6)}\),
\(\frac{57}{28}_{(2)}\), \(\frac{25}{12}_{(8)}\),
\(\frac{43}{20}_{(2)}\),  \\ & \(\frac{61}{28}_{(2)}\),
\(\frac{9}{4}_{(16)}\), \(\frac{47}{20}_{(2)}\),
\(\frac{29}{12}_{(6)}\), \(\frac{5}{2}_{(2)}\), \(\frac{51}{20}_{(2)}\),
\(\frac{11}{4}_{(48)}\), \(\frac{3}{1}_{(2)}\), \(\frac{37}{12}_{(4)}\),
\(\frac{13}{4}_{(8)}\), \(\frac{41}{12}_{(2)}\),
\(\frac{15}{4}_{(26)}\), \(\frac{49}{12}_{(2)}\), \\ & 
\(\frac{17}{4}_{(4)}\), \(\frac{19}{4}_{(16)}\), \(\frac{21}{4}_{(2)}\),
\(\frac{23}{4}_{(12)}\), \(\frac{27}{4}_{(8)}\), \(\frac{31}{4}_{(8)}\),
\(\frac{35}{4}_{(6)}\), \(\frac{39}{4}_{(4)}\),
\(\frac{43}{4}_{(2)}\) \\
\end{longtable}

{We have shown that the Farey sequence $F_N$ and their inverse appear along ${\rm Re}~u=B(k)=0$ when we vary $|e_0|, |e|, |m|,|m_0|=0,\cdots,N$.
The Farey sequence structure would not become manifest when we vary fluxes in a different way, e.g., a region with a tadpole-type quadratic bound on the fluxes.
Note $\beta^2=C(k)/A(k)$ along ${\rm Re}~u=B(k)=0$.
Hence, if there is the number $N_C$ satisfying   $|A(k)|,|C(k)|=1,2,\cdots,N_C$, the Farey sequence $F_{N_C}$ as well as their inverses appears manifestly.
On top of that, for $N_{\rm max}=\max(|A(k)|,|C(k)|)$, some of $F_{N_{\rm max}}$ and their inverses  appear after being thinned out.
}

{Thus, the rational-ratio structure $\beta^2=C(k)/A(k)$ is intrinsic to the vacuum equation on the locus ${\rm Re}~u=B(k)=0$, whereas the appearance of the complete Farey sequence depends on the sampling prescription. In particular, a tadpole-type quadratic bound generally selects only a subset of the pairs $(A(k),C(k))$ and therefore yields a thinned Farey structure rather than the complete sequence produced by the rectangular cutoff.}

\subsubsection{Degeneracy and void}
\label{sec:void}

\begin{figure}
    \centering
    \includegraphics[width=0.75\linewidth]{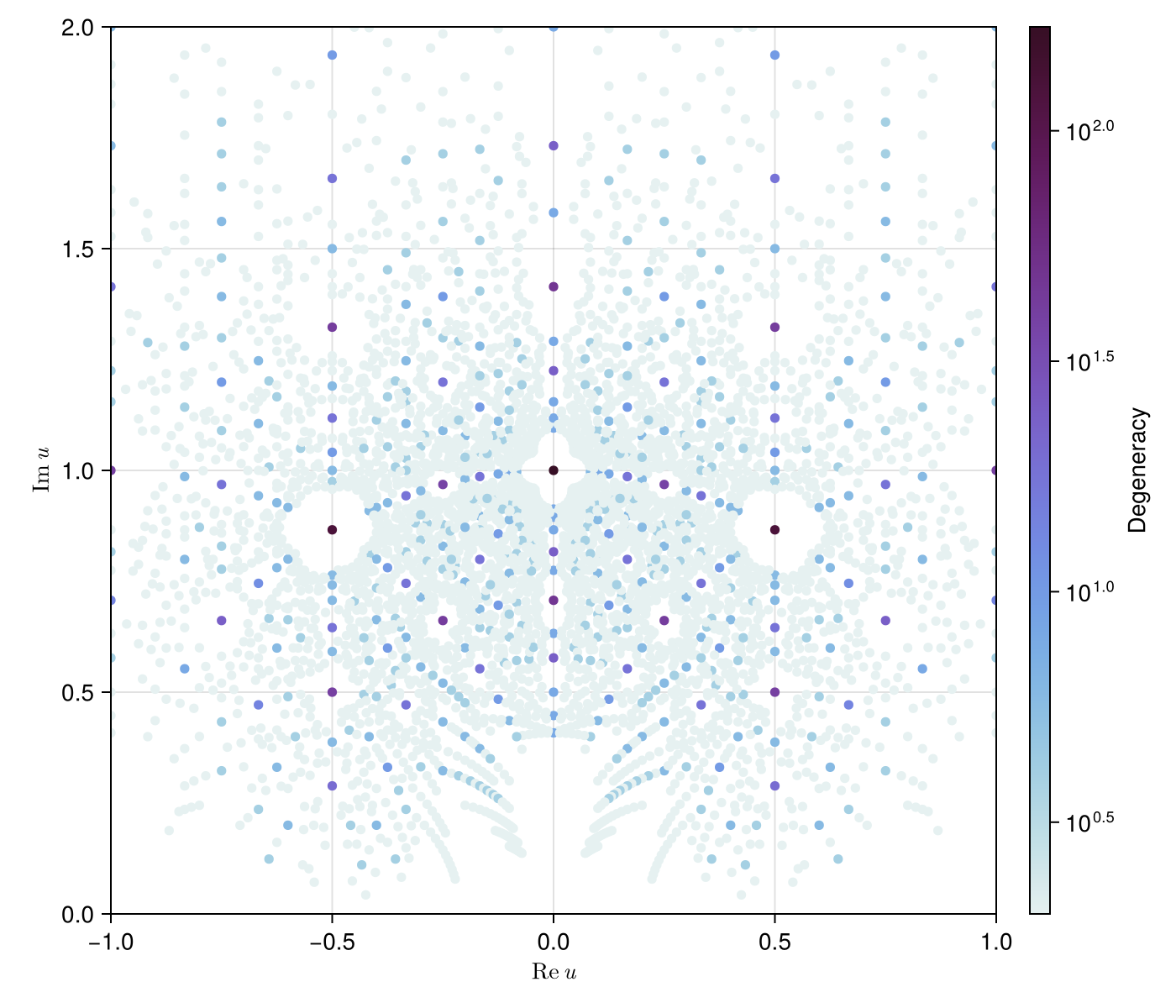}
    \caption{Degeneracies and voids for $N=6$.}
    \label{fig:orientifold_N_6_degeneracy}
\end{figure}

Figures \ref{fig:orientifold_N_12} and \ref{fig:orientifold_N_12_fund} show that there are {hole-like regions containing no solutions for} $u$.
These holes are known as the voids \cite{Denef:2004ze}.
They are located at $u=i$, $e^{\pi i /3}$, and so on.
It is remarkable that these centers of voids have high degeneracies as seen from Tables \ref{tab:beta-orientifold-0} and \ref{tab:beta-orientifold-1/2}.
That suggests that the voids are located at values of $u$ with high degeneracies.
Figure \ref{fig:orientifold_N_6_degeneracy} shows such behavior.
The degeneracy is represented by color. 
Darker colors indicate higher degeneracies. 
The wider void expands at the vacuum with higher degeneracy.

\begin{figure}
    \centering
    \includegraphics[width=0.75\linewidth]{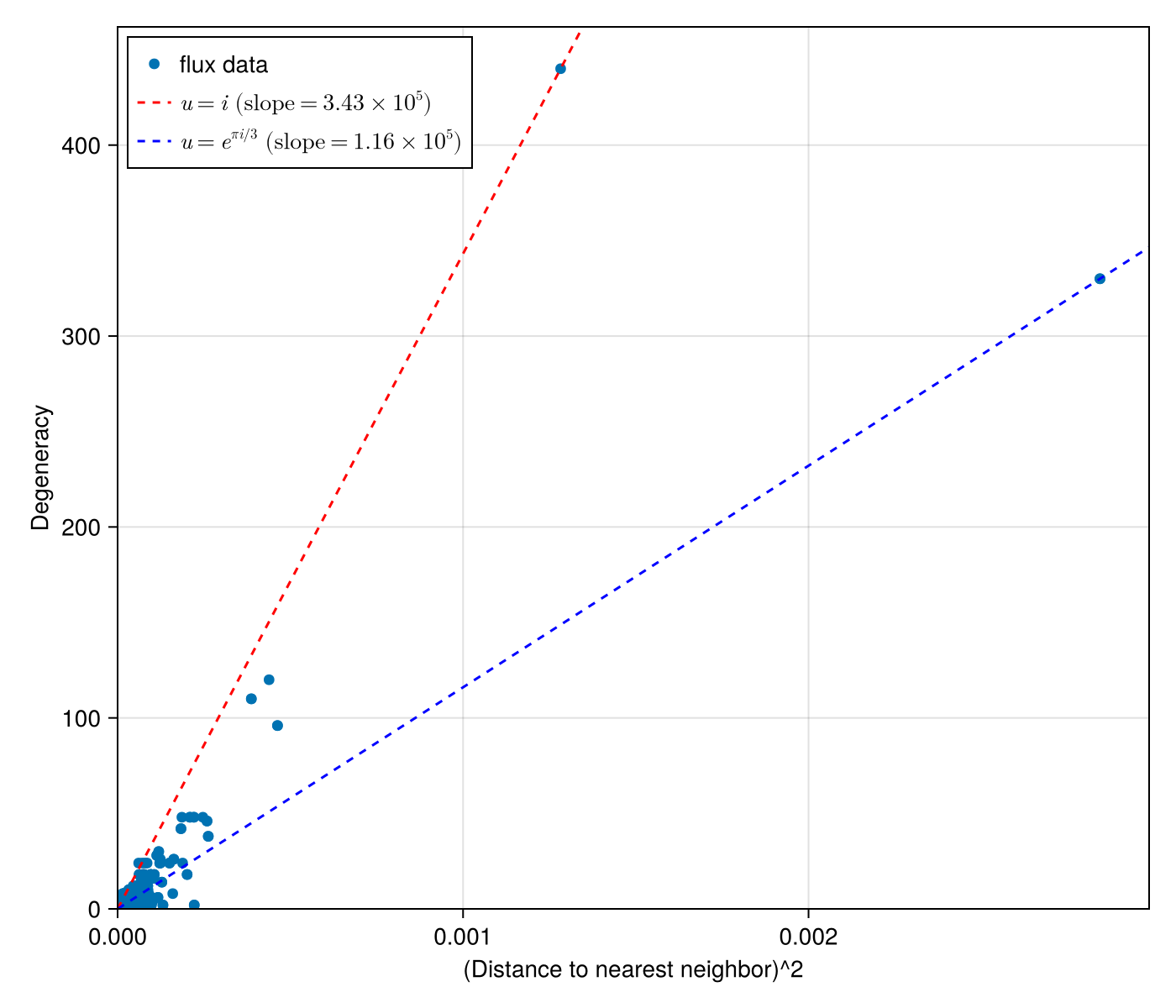}
    \caption{Correlation between the degeneracy and the void area for $N=10$.}
    \label{fig:degeneracy_N_10}
\end{figure}

\begin{figure}
    \centering
    \includegraphics[width=0.75\linewidth]{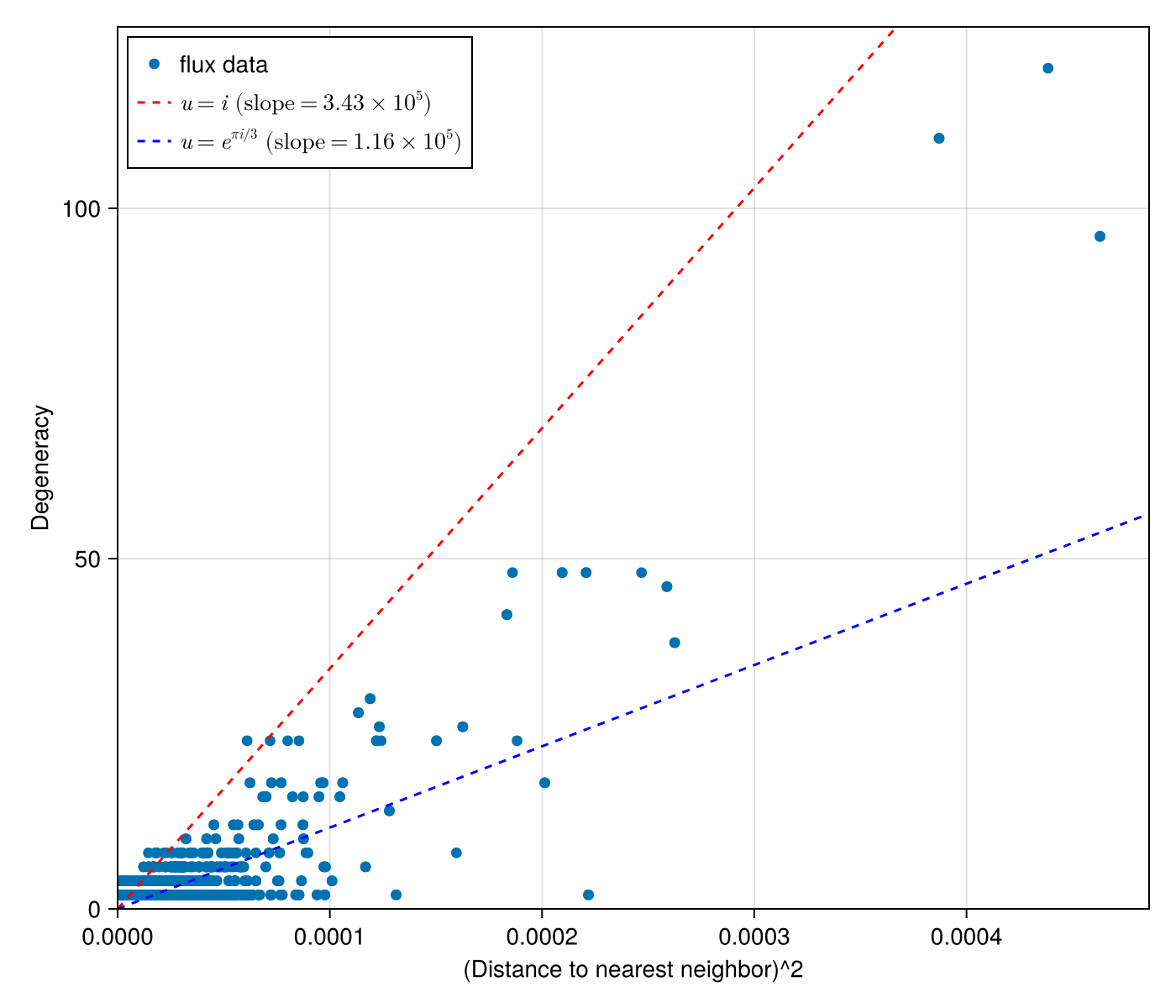}
    \caption{Enlarged view of Figure \ref{fig:degeneracy_N_10}.}
    \label{fig:degeneracy_N_10-2}
\end{figure}

\begin{figure}
    \centering
    \includegraphics[width=0.75\linewidth]{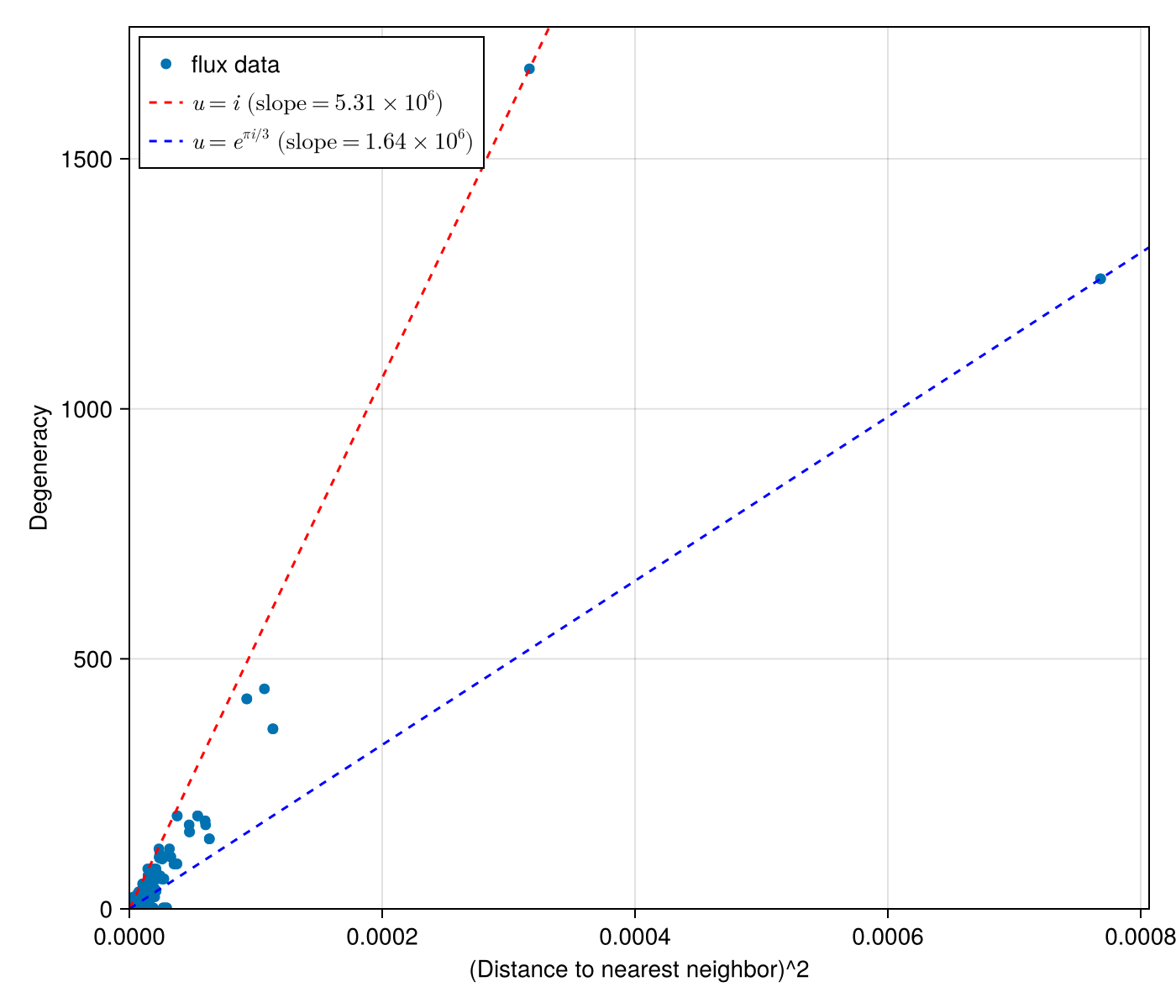}
    \caption{Correlation between the degeneracy and the void area for $N=20$.}
    \label{fig:degeneracy_N_20}
\end{figure}

\begin{figure}
    \centering
    \includegraphics[width=0.75\linewidth]{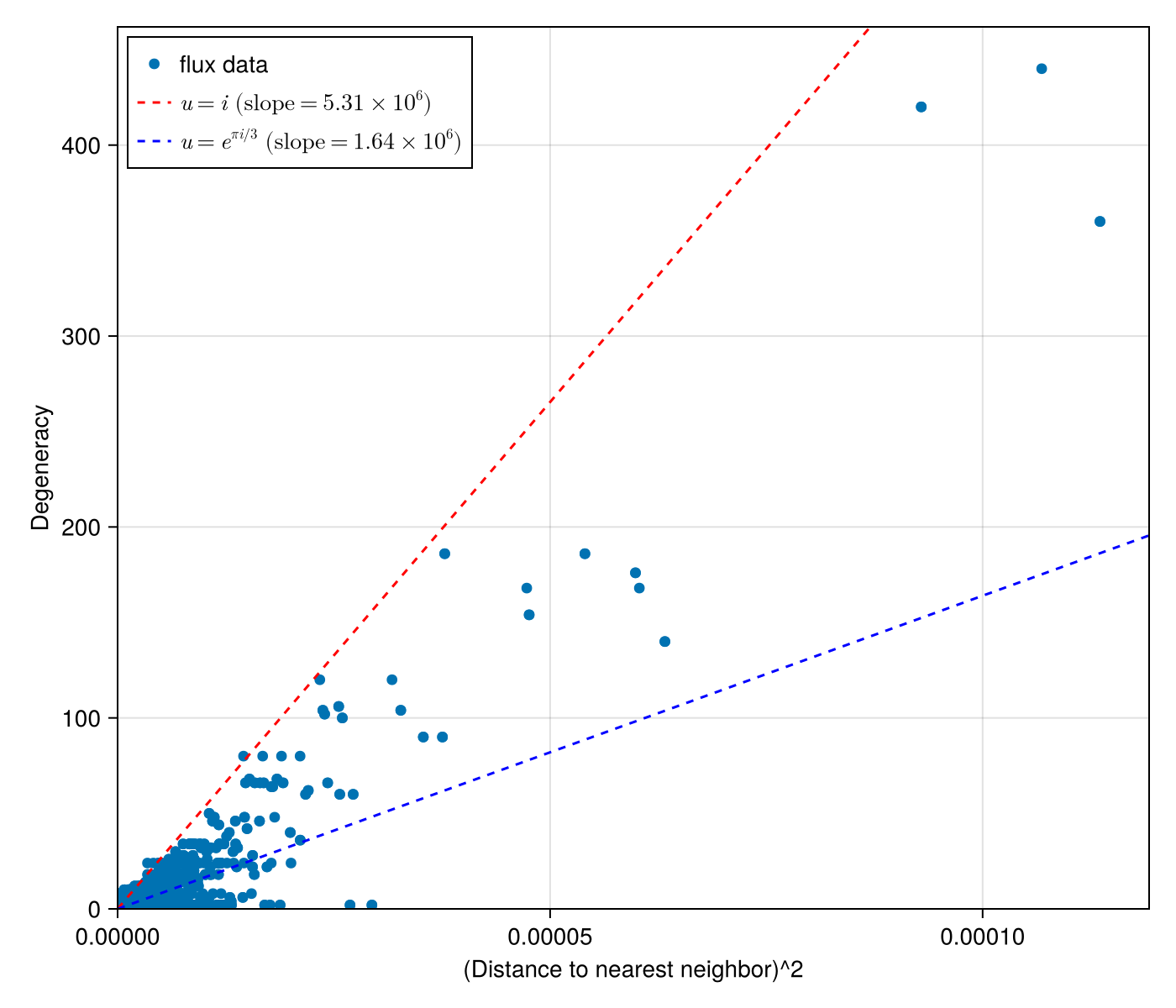}
    \caption{Enlarged view of Figure \ref{fig:degeneracy_N_20}.}
    \label{fig:degeneracy_N_20-2}
\end{figure}

\begin{figure}
    \centering
    \includegraphics[width=0.75\linewidth]{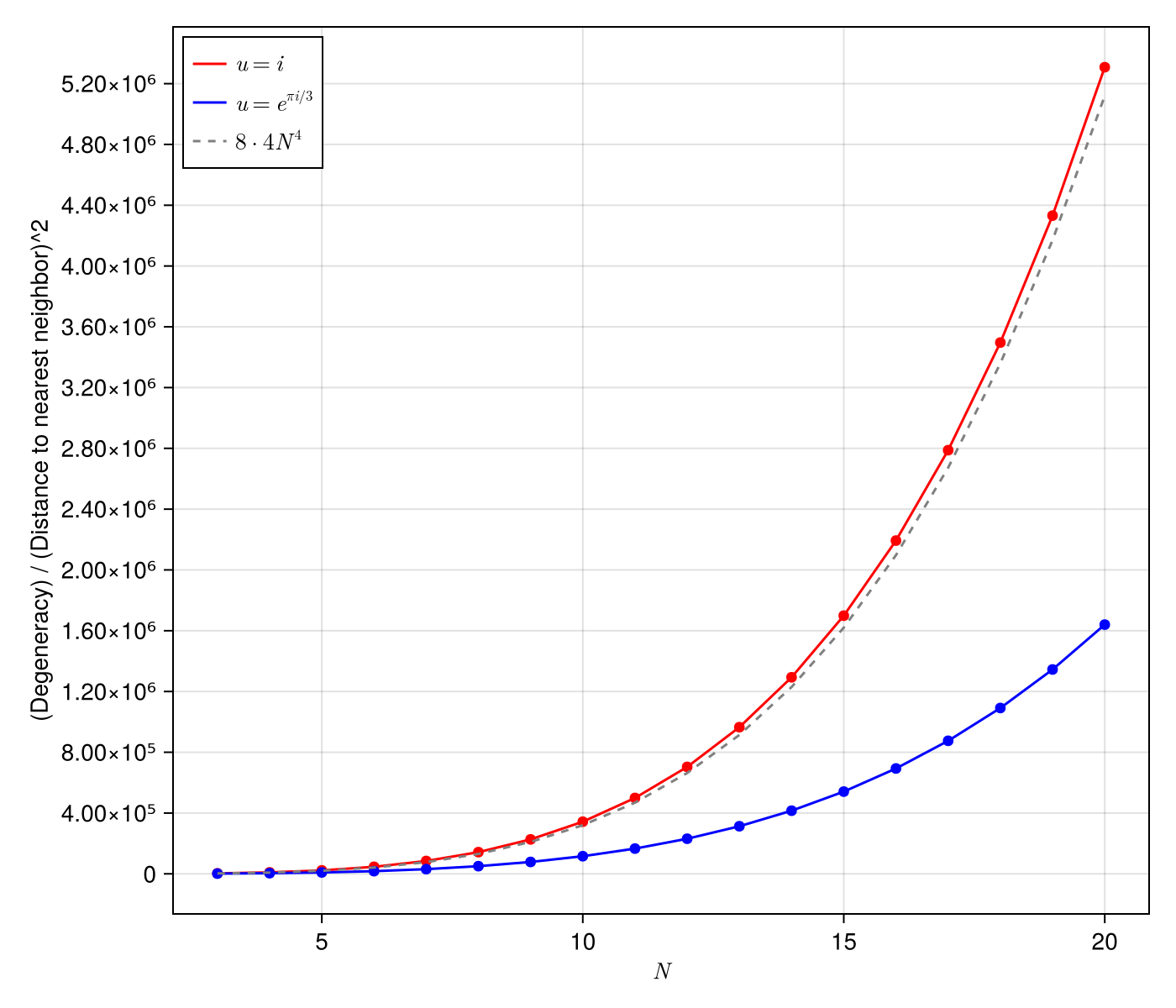}
    \caption{$N$ dependence of slopes of red and blues lines in Figures \ref{fig:degeneracy_N_10}-\ref{fig:degeneracy_N_20-2}. 
    The dotted curve corresponds to $8\cdot 4N^4$.}
    \label{fig:degeneracy_slope_N}
\end{figure}

Let us assume that vacua are distributed roughly evenly.
However, there are values of $u$ that can be realized with high degeneracies.
They ``absorb'' the surrounding vacua, which become voids.
The area of void may correlate with the degeneracy of vacuum at the center.
Figures \ref{fig:degeneracy_N_10} and \ref{fig:degeneracy_N_10-2} show correlations {for $N=10$}.
The red (blue) line connects the origin and $u=i$ ($u=e^{\pi i/3}$).
The rough linear relation is found.
{Figures \ref{fig:degeneracy_N_20} and \ref{fig:degeneracy_N_20-2} show similar behavior for $N=20$.
Thus, we can find the rough linear relation for fixed $N$.
However, Figures \ref{fig:degeneracy_N_10}-\ref{fig:degeneracy_N_20-2} show that the slopes depend on $N$.
Figure \ref{fig:degeneracy_slope_N} shows such behavior.
This behavior can be understood analytically as follows.
We examine the behavior at $u=i$.
For $N$, the nearest neighbor point along ${\rm Re}~u=0$ is $u=i(1-1/N)$ because of the Farey sequence structure.
Their distance and its square are $1/N$ and $1/N^2$.
The degeneracy at $u=i$ is equal to $(2N+1)^2-1$ as evaluated in the previous section.
Thus, the ratio of the degeneracy to the distance squared to the nearest neighbor along ${\rm Re}~u=0$ at $u=i$ is estimated by 
\begin{align}
    \frac{(2N+1)^2-1}{N^{-2}} \sim 4 N^4.
\end{align}
This estimation explains roughly the behavior of the red curve in Figure \ref{fig:degeneracy_slope_N}.
Indeed, $u=i(1-1/N)$ is not nearest point to $u=i$, but there is the nearest point along the direction ${\rm Re}~u\neq 0$.
For example, the function $8\cdot 4N^4$ fits well the red curve in Figure \ref{fig:degeneracy_slope_N}.
}

{The box scan does not give a uniform sampling of moduli space. For each fixed $N$, it shows a correlation between the degeneracy and the void size, but the slope changes with $N$. The detailed void structure may also change if a different flux cutoff or sampling method is used.}

\begin{figure}
    \centering
    \includegraphics[width=0.75\linewidth]{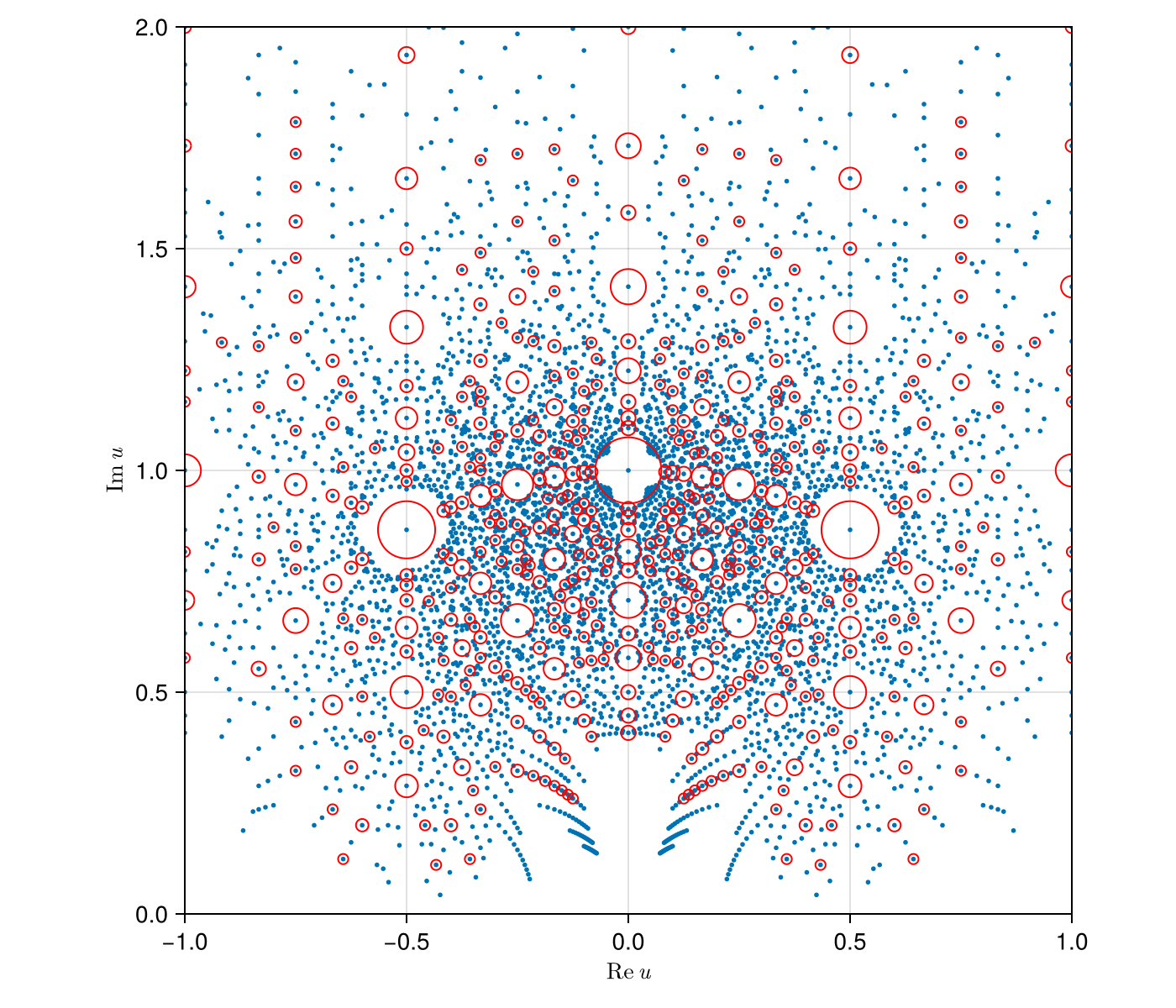}
    \caption{Degeneracy and void for $N=6$. Red circles with $(\text{radius}) = \sqrt{\text{degeneracy}}/174$ are drawn.}
    \label{fig:void-1}
\end{figure}

\begin{figure}
    \centering
    \includegraphics[width=0.75\linewidth]{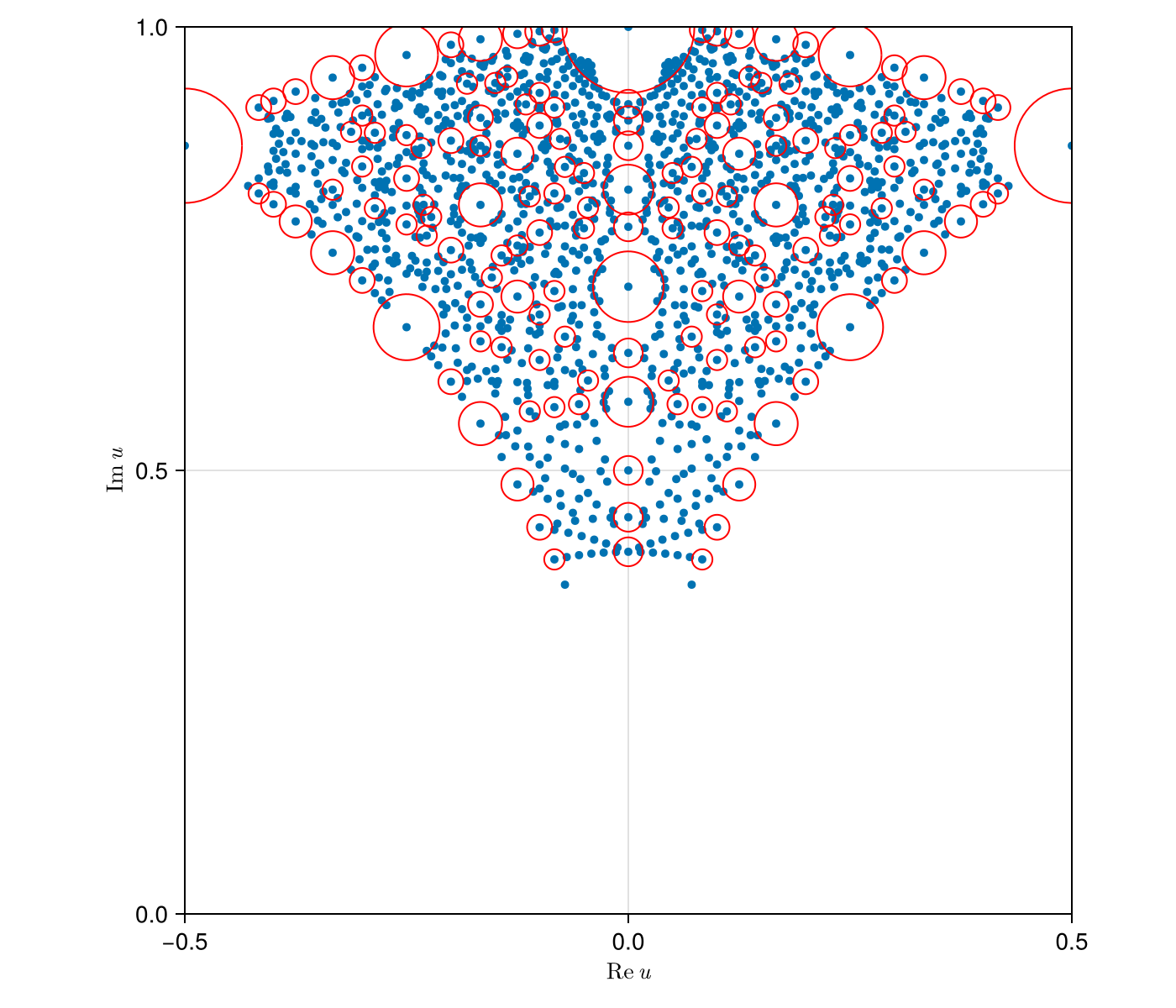}
    \caption{Degeneracy and void for $N=6$. Red circles with $(\text{radius}) = \sqrt{\text{degeneracy}}/174$ are drawn. }
    \label{fig:void-2}
\end{figure}

The average of slopes of red and blue lines is equal to 174 {when $N=6$}.
Figure \ref{fig:void-1} corresponds to Figure \ref{fig:orientifold_N_12} for $N=6$.
We have drawn additionally red circles, whose radii are equal to $\sqrt{d}/174$, where $d$ is degeneracy.
The region $|u|<1$ is zoomed up in 
Figure \ref{fig:void-2}.

\subsection{Relations among vacua by modular symmetry}
\label{sec:relation}

Here, we study relations among vacua.
Each vacuum value $u$ is obtained as a solution of the holomorphic quadratic equation $F(u;k)=0$ in Eq.~(\ref{eq:moduli_quadratic_eq}).
Thus, we examine symmetries of the equation in order to study relations among vacua.
Coefficients $A(k), B(k), C(k)$ are functions of fluxes.
For simplicity, we consider their integer values $A, B, C$ here, i.e.,
\begin{align}\label{eq:F=0}
    F(u)=Au^2+Bu+C=0.
\end{align}

Here, we study the modular transformation behavior of the above equation $F(u)=0$.
By the $S$ transformation, $u \to -1/u$, the equation becomes $F(-1/u)=0$, where
\begin{align}
    F(-\frac{1}{u})=\frac{1}{u^2}(Cu^2-Bu+A).
\end{align}
Also, by the {$T^{-1}$} transformation, {$u \to u-1$}, the equation becomes {$F(u-1)=0$},
where
\begin{align}
    F(u+1)=Au^2+ (2A+B)u+(A+B+C).
\end{align}
Here, we define modular transformation of coefficients by
\begin{align}
    (\gamma  F)(u)=(cu+d)^2F(\gamma^{-1} u),
\end{align}
where $\gamma \in SL(2,\mathbb{Z})$ and 
$(cu+d)^2$ is automorphy factor of weight 2.
Then coefficients transform as
\begin{align}
    &S: \pmqty{A \\ B \\ C}
        \longrightarrow
        \pmqty{C \\ -B \\ A}
        = \pmqty{0&0& 1 \\ 0& -1 & 0\\ 1 &0&0} \pmqty{A \\ B \\ C}, \notag \\
    &T: \pmqty{A \\ B \\ C}
        \longrightarrow
        \pmqty{A \\ -2A + B \\ A -B + C}
        = \pmqty{1 &0&0 \\ -2 & 1 & 0\\ 1 & -1 & 1} \pmqty{A \\ B \\ C}.
\end{align}
They satisfy $S^2=(ST)^3=I$, i.e., $PSL(2,\mathbb{Z})$.
We identify combinations of coefficients $F=(A,B,C)$ by $PSL(2,\mathbb{Z})$, and  
define the equivalence classes $[F] = [A,B,C] \in \mathbb{Z}^3/\mathrm{PSL}(2,\mathbb{Z})$.
When we choose a proper representative in $[F]$, the solution of $F(u)=0$ corresponds to a single point in the fundamental domain $\mathcal{F}$,
\begin{equation}
    [F] \xrightarrow{\text{Solution of $[F](u) = 0$}} \tau \in \mathcal{F}.
\end{equation}

It is found that both the discriminant $D$
\begin{align}
    D=B^2-4AC,
\end{align}
and the greatest common divisor of $(A,B,C)$,
\begin{align}
    n={\gcd}(A,B,C),
\end{align}
are invariant under $PSL(2,\mathbb{Z})$.
Therefore, a solution of $F(u)=0$ is specified by $D$ and $n$, and such a vacuum can be represented by 
\begin{equation}
    \ket{[A,B,C]} = \ket{D,n,\alpha}.
\end{equation}
In general, only $D$ and $n$ are not enough to specify a single equivalence class $[A,B,C]$, and we need additional quantity $\alpha$.

\subsection{Relations among vacua}
\label{sec:relation-2}

Here, we study relations among vacua other than the modular symmetry.

\subsubsection{Multiplications of coefficients}
\label{sec:scaling}

We multiply natural number $k$ by coefficients,
\begin{align}
    &k: \pmqty{A \\ B \\ C}
        \longrightarrow
        \pmqty{kA \\ kB \\ kC}
        = \pmqty{k && \\ & k & \\ && k} \pmqty{A \\ B \\ C}, \quad k \in \mathbb{N}.
\end{align}
That changes the equations 
\begin{align}
    F(u)=0 \to kF(u)=0.
\end{align}
The solution never changes, but the discriminant $D$ and gcd $n$ change as
\begin{align}
    D\to k^2D, \qquad n \to kn.
\end{align}
Note that if $\gcd(A,B,C)=n$, $D/n^2$ is integer.
Thus, the equivalence classes $[A,B,C]$ with $n=1$ are important.
We focus on $n=1$ in what follows.

\subsubsection{Scaling of modulus}
\label{sec:scaling-modulus}

We scale the coefficients by natural integer $k$ as follows,
\begin{align}
    &\omega_k: \pmqty{A \\ B \\ C}
        \longrightarrow
        \pmqty{k^2A \\ kB \\ C}
        = \pmqty{k^2 && \\ & k & \\ && 1} \pmqty{A \\ B \\ C}, \quad k \in \mathbb{N}.
\end{align}

{Under this scaling, }$D$ changes as $D \to k^2D$.
The equation (\ref{eq:F=0}) becomes,
\begin{align}
    k^2Au^2+kBu+C=F(ku)=0.
\end{align}
Thus, the above scaling changes the solution as
\begin{align}
    \omega_k:u\to \frac{u}{k}.
\end{align}

\subsubsection{Abelian symmetry by Gauss composition}
\label{sec:gauss}

Let us consider two equivalence classes $[F_1]$ and $[F_2]$.
In general, they are different from each other, even if their discriminant $D$ and gcd $n$ are the same.
For example, the following equivalence classes:
\begin{equation}
    [F_1] = [1, 0, 5], \quad [F_2] = [2, 2, 3]
\end{equation}
have $D=-20$ and $n=1$, but they are different from each other.
Indeed, they correspond to 
\begin{align}
    u=i\sqrt{5}, \quad u=\frac{-1+i\sqrt{5}}{2}.
\end{align}

They can be related by Gauss composition method as follows.
The Gauss composition is the method to relate quadratic equations with the same discriminant $D$.
Let us consider the following two equivalence classes and their representatives:
\begin{equation}
    [F_1] = [A_1, B, C_1], \quad
    [F_2] = [A_2, B, C_2],  
\end{equation}
which have the same discriminant $D$ and gcd $n=1$, where $B$ is the same, and $A_1$ and $A_2$ are coprime, i.e.,  $\gcd(A_1, A_2) = 1$.
Then we define the following multiplication:
\begin{equation}
    [F_1] * [F_2] = \left[{A_1A_2, B, \frac{B^2-D}{4A_1A_2}}\right].
\end{equation}
That is the Gauss composition law.
Note that the discriminant $D$ and gcd $n=1$ remain unchanged by the Gauss composition.

For example, we can compute 
\begin{gather}
        [1,0,5]* [1,0,5] = [1,0,5], \\
    [2,2,3] * [1,0,5] = [1,0,5] * [2,2,3] = [2,2,3], \\
    [2,2,3]*[2,2,3]  = [1,0,5].
\end{gather}
These algebraic relations are isomorphic to the $\mathbb{Z}_2$ symmetry.
We denote its elements by $e,g$, which satisfy 
$e^2=e$, $eg=ge=g$, and $g^2=e$.
We can write 
\begin{align}
     \ket{[F_1]}&= \ket{[1,0,5]} = \ket{D=-20,n=1,e}, \notag \\
     \ket{[F_2]}&= \ket{[2,2,3]} = \ket{D=-20,n=1,g}.
\end{align}

As shown by the above examples, vacuum labeled by $[F_i]$ are related by the Gauss composition law, and their symmetries are Abelian like $\mathbb{Z}_N\times \mathbb{Z}_{N'}\times \cdots$.
The identities for generic $D$ and $n=\gcd(A,B,C)=1$ can be written by
\begin{align}
    e = 
      \left\{ \,
      \begin{aligned}
        & \Big[1,0,-\frac{D}{4}\Big],\quad D\equiv0\pmod4 \\
        & \Big[1,1,\frac{1-D}{4}\Big],\quad D\equiv1\pmod4\\
      \end{aligned}.
      \right.
\end{align}
For example, for $D=-20$, we find $e=[1,0,5]$, which is the above one.

The inverse of $[A,B,C]$ is obtained as 
\begin{align}
    [A,B,C ]^{-1} = [A,-B,C].
\end{align}
The modulus value corresponding to $[F]=[A,B,C]$ is written by
\begin{align}
    u_F = \frac{-B+i\sqrt{-D}}{2A}.
\end{align}
The inverse $[F]^{-1} =[A,-B,C]$ corresponds to 
\begin{align}
    u_{F^{-1}} =\frac{B+i\sqrt{-D}}{2A} = -\bar{u}_F.
\end{align}
That is the CP transformation,
\begin{align}
\label{eq:CP}
    u \to -\bar u.
\end{align}
The points on the line $u=i\beta$ corresponding to $[A,0,C]$ are CP-symmetric, and the equivalence classes $[A,0,C]$ are self-conjugate,
\begin{align}
    [A,0,C]*[A,0,C]= e,
\end{align}
where $e$ denotes the identity of $\mathbb{Z}_N$, which is constructed by the Gauss composition.
If $N$=odd, the class $[A,0,C]$ is the identity $e$.
If $N=$ even, the class $[A,0,C]$ is $e$ or $g$, where 
$g$ is the $\mathbb{Z}_2$ element in $\mathbb{Z}_{2N}$, i.e., $g^2=e$ and $g\neq e$.

Similarly, the points on the line ${\rm Re }~u=\pm1/2$ are CP-symmetric up to $T$ transformation and the corresponding equivalence classes $[A,\pm A,C]$ are self-conjugate, 
\begin{align}
    [A,\pm A,C]*[A,\pm A,C]= e.
\end{align}
This class is $e$ when $N=$ odd, and $e$ or $g$ when $N=$ even.
Furthermore, the points on the arc $|u|=1$ are CP-symmetric up to the $S$ transformation.
That is because the $S$ transform $u=e^{i\theta}$ to $-1/u=-e^{-i\theta}$, which is nothing but the CP conjugate $-\bar u$.
These points correspond to the classes $[A,B,A]$.
Again, they satisfy 
\begin{align}
    [A,B,A]*[A,B,A]=e.
\end{align}
Thus, the Abelian symmetry such as $\mathbb{Z}_N \times \mathbb{Z}_{N'}\times \cdots $ symmetry generated by the Gauss composition includes the CP symmetry.

\begin{figure}
\centering
\includegraphics[width=0.75\linewidth,height=\textheight,keepaspectratio]{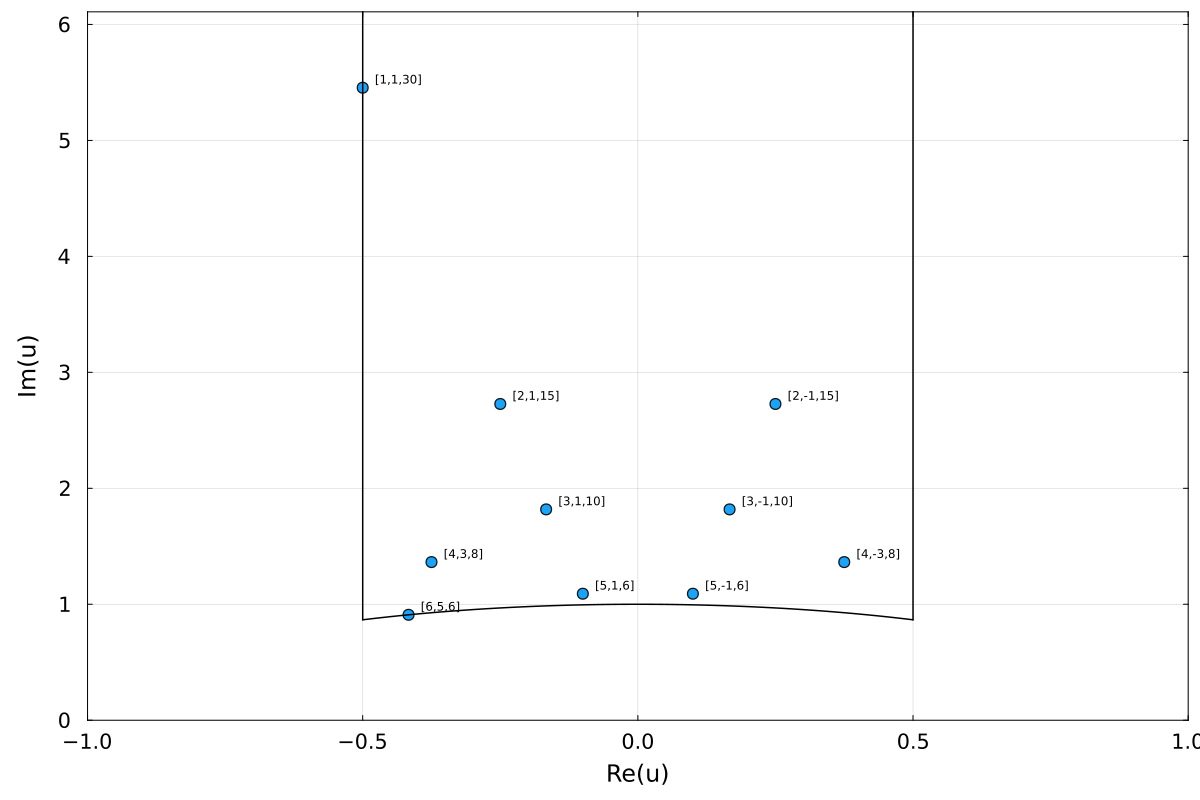}
\caption{$\mathbb{Z}_{10}$ symmetric vacua for $D=-119$.}
\label{fig:Z10}
\end{figure}

\begin{figure}
\centering
\includegraphics[width=0.75\linewidth,height=\textheight,keepaspectratio]{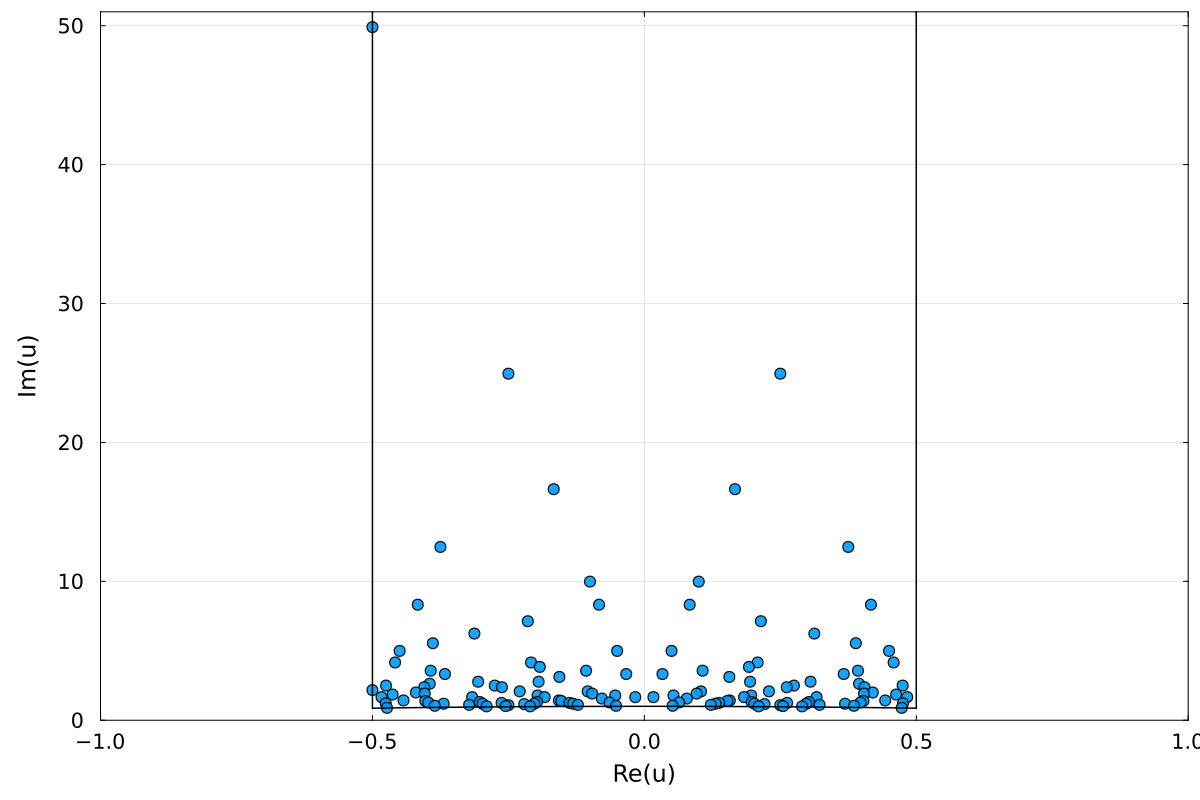}
\caption{$\mathbb{Z}_{130}$ symmetric vacua for $D=-9959$.}
\label{fig:Z130}
\end{figure}

Table~\ref{tab:gauss} shows examples of vacua related by the Gauss compositions for smaller $|D|$.
For smaller $|D|$, the order is smaller like $N=1,2,3$.
The fixed points $u=i$ and $e^{\pi i/3}$ are singlets.
Figure \ref{fig:Z10} shows $\mathbb{Z}_{10}$ symmetric vacua for $D=-119$.
Figure \ref{fig:Z130} shows $\mathbb{Z}_{130}$ symmetric vacua for $D=-9959$.
Therefore, these vacua are specified by $\mathbb{Z}_{10}$ charges and $\mathbb{Z}_{130}$ charges, respectively, in addition to $D$ and $n=1$.
These figures are symmetric under the CP transformation.
Note that ${\rm u}=0, \pm 1/2$ and $|u|=1$ are CP symmetric.
Furthermore, these figures have specific patterns.

\begin{longtable}{ccl}
    \caption{$\mathbb{Z}_N$ symmetric vacua through the Gauss compositions.}
    \label{tab:gauss}\\ \hline
    $D$ & order of $\mathbb{Z}_N$ & $[A,B,C]$ in fundamental domain \\ \hline
    \endfirsthead
    \hline
    $D$ &   order of $\mathbb{Z}_N$ & $[A,B,C]$ in fundamental domain \\ \hline
    \endhead
    \hline
    \endfoot
    \hline
    \endlastfoot
\(-3\) & 1 & \([1, 1, 1]\)  \\
\(-4\) & 1 & \([1, 0, 1]\)  \\
\(-7\) & 1 & \([1, 1, 2]\)  \\
\(-8\) & 1 & \([1, 0, 2]\)  \\
\(-11\) & 1 & \([1, 1, 3]\) \\
\(-12\) & 1 & \([1, 0, 3]\)  \\
\(-15\) & 2 & \([1, 1, 4], [2, 1, 2]\) \\
\(-16\) & 1 & \([1, 0, 4]\) \\
\(-19\) & 1 & \([1, 1, 5]\)  \\
\(-20\) & 2 & \([1, 0, 5], [2, 2, 3]\) \\
\(-23\) & 3 & \([1, 1, 6], [2, -1, 3], [2, 1, 3]\) \\
\(-24\) & 2 & \([1, 0, 6], [2, 0, 3]\)  \\
\(-27\) & 1 & \([1, 1, 7]\)  \\
\(-28\) & 1 & \([1, 0, 7]\)  \\
\(-32\) & 2 & \([1, 0, 8], [3, 2, 3]\)\\
\(-35\) & 2 & \([1, 1, 9], [3, 1, 3]\)\\
\(-36\) & 2 & \([1, 0, 9], [2, 2, 5]\) \\
\(-39\) & 4 & \([1, 1, 10], [2, -1, 5], [2, 1, 5], [3, 3, 4]\) \\
\(-40\) & 2 & \([1, 0, 10], [2, 0, 5]\) \\
\end{longtable}

\newpage

\subsection{CP violation}

As mentioned in the previous subsection, the CP symmetry transforms the modulus
as Eq.~(\ref{eq:CP}),
up to the modular transformation.
The modulus stabilization can break the CP symmetry.
The CP symmetry remains  at ${\rm Re}~   u=\alpha=0,\pm 1/2$ and $|u|=1$.
The CP symmetry is violated at other vacua.
The CP symmetric vacua seem to be preferable to the CP violating vacua.
{The generic} superpotential in Eq.~(\ref{eq:W-orientifold}) is not CP symmetric, although {the} K\"ahler potential is.
The CP invariance requires that $|W|^2$ is invariant under the transformation (\ref{eq:CP}).
There are two possibilities.
The first one is $m_0=e=0$, while the second one is $m=e_0=0$.
The first one leads to the following SUSY condition:
\begin{align}
    F(u)=-m^2u^2+e_0m=0,
\end{align}
and the second one leads the following condition:
\begin{align}
    F(u)=em_0u^2-e^2=0.
\end{align}
We require ${\rm Im}~u>0$ as a physical condition in both cases.
Then we find that the solution corresponds to 
${\rm Re}~u=0$ in both.
Indeed, the above quadratic equations have $B=0$.
The discussion on the equivalence classes and CP-symmetry also leads to the same result.
That is CP invariant vacuum.
That is, when we start with the CP invariant superpotential, the potential minimum is CP symmetric.
One cannot realize spontaneous CP violation \cite{Kobayashi:2020uaj,Ishiguro:2020nuf}.

\section{Calabi-Yau compactification}
\label{sec:CY}

Similarly, we can study Calabi-Yau compactifications.
For simplicity, we consider the model with a single modulus.
As an illustrative example, we consider the mirror manifold of a complete intersection Calabi-Yau threefold  with $h^{1,1}=1$, whose K\"ahler potential and superpotential induced by RR fluxes are written by 
{
\begin{align}\label{eq:cicy_def}
    K &= - \ln \kappa, \quad \kappa \equiv \frac{{i} \kappa_{111}}{6} (u - \bar{u})^3, \notag \\ W &= \kappa_{111}\left( e_0 + e u + \frac{1}{2} m u^2 + \frac{1}{6} m_0 u^3 \right).
\end{align}
There are several complete intersection Calabi-Yau (CICY) threefolds with $h^{1,1}=1$ and they have various intersection numbers, e.g., $\kappa_{111}=5,8,9,12,16$ \cite{CICY}.
However, the value of $\kappa_{111}$ is irrelevant to solve the F-term condition (\ref{eq:SUSY-min}) because $\kappa_{111}$ can be factored out in Eq.~(\ref{eq:SUSY-min}).
In what follows, we set $\kappa_{111}=1$. 
}
{More explicitly, the examples underlying Eq.~(\ref{eq:cicy_def}) are the mirrors of the one-parameter CICYs:}
\begin{align}
&{X_5\subset\mathbb{P}^4,\qquad
X_{4,2}\subset\mathbb{P}^5,\qquad
X_{3,3}\subset\mathbb{P}^5,}\notag\\
&{X_{3,2,2}\subset\mathbb{P}^6,\qquad
X_{2,2,2,2}\subset\mathbb{P}^7,}
\end{align}
{whose triple intersection numbers are $\kappa_{111}=5,8,9,12,16$, respectively.}

{Equation~(\ref{eq:cicy_def}) is the leading large-complex-structure approximation. We retain only the cubic term and neglect the remaining polynomial terms in the prepotential, the perturbative $\zeta(3)$ contribution, and world-sheet instanton corrections. The instanton corrections are suppressed as $e^{-2\pi {\rm Im}\,u}$, while the relative perturbative correction is suppressed by inverse powers of ${\rm Im}\,u$.}
{Not all solutions in the flux scan lie deep in the large-complex-structure region. The figures show the arithmetic structure obtained in the classical cubic approximation. Only solutions with sufficiently large ${\rm Im}\,u$ are reliably controlled. Points near the boundary require the full period vector, and we leave their analysis for future work.}

\subsection{Gr\"obner basis}

The Gr\"obner cover is obtained as 
\begin{equation}\phantomsection\label{eq:cicy_groebner}{
\begin{cases}
  4\beta^2e^2m_0^2 - 4\beta^2em^2m_0 + \beta^2m^4 + 9e_0^2m_0^2 - 18e_0emm_0 + 6e_0m^3 + 8e^3m_0 - 3e^2m^2=0, \\
  2\alpha em_0 - \alpha m^2 + 3e_0m_0 - em=0, \\
  3\alpha e_0m_0 - \alpha em - 2\beta^2em_0 + \beta^2m^2 + 6e_0m - 4e^2=0, \\
  3\alpha e_0m^2 - 2\alpha e^2m - 4\beta^2e^2m_0 + 2\beta^2em^2 - 9e_0^2m_0 + 15e_0em - 8e^3=0, \\
  \alpha^2m_0 + 2\alpha m + \beta^2m_0 + 2e=0, \\
  \alpha^2m + 4\alpha e + \beta^2m + 6e_0=0, \\
\end{cases}
}\end{equation} 
under the condition $m^2-2em_0 \neq 0$.

The first and second equations mean that the solution $u=\alpha+i\beta$ of the SUSY condition is written by 
\begin{equation}\label{eq:cicy_alpha}
\alpha = \frac{-(6e_0m_0-2em)}{2(2em_0-m^2)},
\end{equation} \begin{equation}\label{eq:cicy_beta}
\beta = \sqrt{\frac{4(2em_0-m^2)(6e_0m-4e^2)-(6e_0m_0-2em)^2}{4(2em_0-m^2)^2}}.
\end{equation} 
The third and fourth equations are satisfied if the above equations (\ref{eq:cicy_alpha}) and (\ref{eq:cicy_beta}) are satisfied.

The fifth and sixth equations in Eq.~(\ref{eq:cicy_groebner}) are written by
\begin{equation}\label{eq:cicy_circ_m_m0}
{\alpha + \frac{m}{m_0}}^2 + \beta^2 = -\frac{2em_0-m^2}{m_0^2}, 
\end{equation} \begin{equation}\label{eq:cicy_circ_e_m}
{\alpha + \frac{2e}{m}}^2 + \beta^2 = -\frac{6e_0m-4e^2}{m^2}.
\end{equation} 
These represent circles.
These equations are also satisfied if the above equations (\ref{eq:cicy_alpha}) and (\ref{eq:cicy_beta}) are satisfied.

\begin{figure}
\centering
\includegraphics[width=0.6\linewidth,height=\textheight,keepaspectratio]{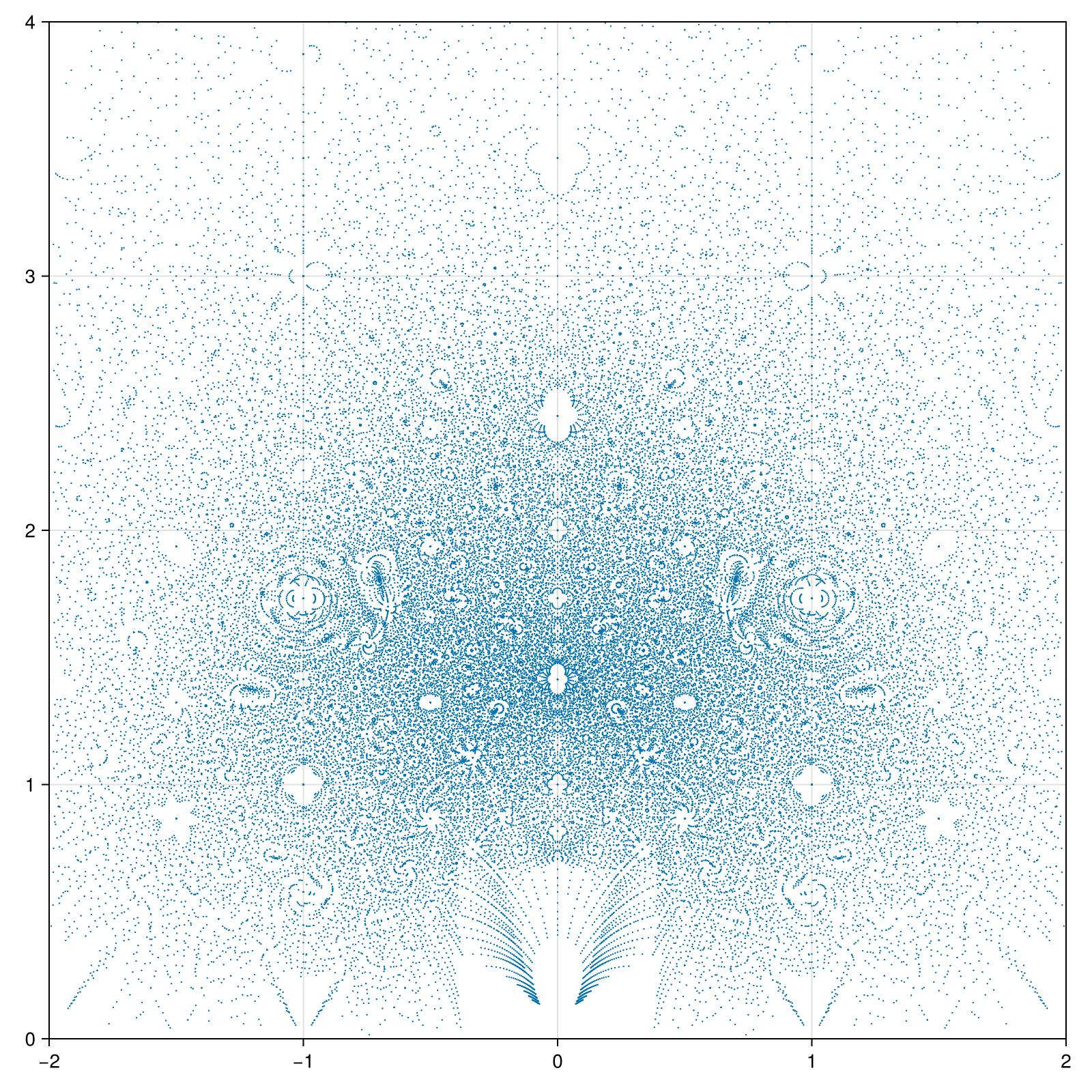}
\caption{Solutions of the SUSY conditions 
(\ref{eq:cicy_alpha}, \ref{eq:cicy_beta}) in Calabi-Yau compactification for 
\(-12 \leq e_0, e, m, m_0 \leq 12\).}\label{fig:cicy_N_12}
\end{figure}

\begin{figure}
\centering
\includegraphics[width=0.6\linewidth,height=\textheight,keepaspectratio]{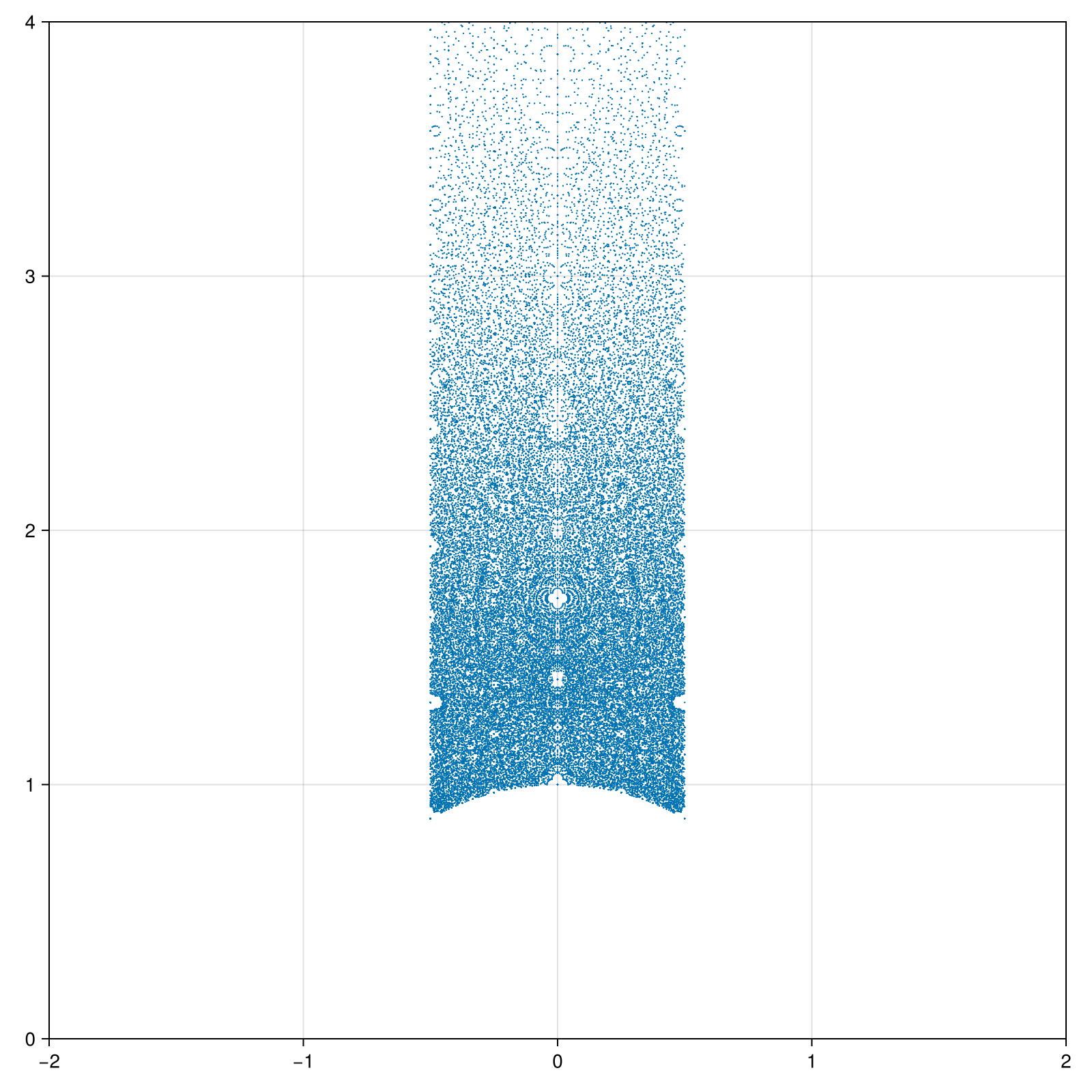}
\caption{Solutions of the SUSY conditions 
(\ref{eq:cicy_alpha}, \ref{eq:cicy_beta}) in the fundamental domain.}\label{fig:cicy_N_12_fund}
\end{figure}

We vary the fluxes in the region \(-N \leq e_0, e, m, m_0 \leq N \) and analyze the SUSY condition.
Figure \ref{fig:cicy_N_12} shows the solution of $u$ for $N=12$ in the upper half plane. 
In Figure \ref{fig:cicy_N_12_fund}, we use the modular transformation such that the point outside the fundamental domain are transformed to the inside.
These patterns are different from those for $T^6/(\mathbb{Z}_2\times\mathbb{Z}_2')$.

Combining Eqs.~(\ref{eq:cicy_alpha}) and (\ref{eq:cicy_beta}), the solution $u $ is written by Eq.~(\ref{eq:u-sol}), where 
\begin{equation}
\begin{aligned}
  A(k)&=2em_0-m^2, \quad  B(k)&=6e_0m_0-2em, \quad C(k)&=6e_0m-4e^2. \\
\end{aligned}
\end{equation}
That is, the solution $u$ satisfies the holomorphic quadratic equation (\ref{eq:moduli_quadratic_eq}).
This equation can be derived by other ways as in the previous section.

The holomorphic quadratic equation, $F(u;k)=0$, in Eq.~(\ref{eq:moduli_quadratic_eq}) was important 
in the analyses of the previous section on 
the $T^6/(\mathbb{Z}_2\times\mathbb{Z}_2')$.
We can study vacuum properties of this Calabi-Yau compactification by analyzing the holomorphic quadratic equation $F(u;k)=0$ by 
replacing 
\begin{align}
    e \to 2e, \quad e_0 \to 6e_0, \quad m_0 \to m_0, \quad m\to m,
\end{align}
in the coefficients, $A(k)$, $B(k)$, and $C(k)$.

\subsection{Modular symmetry}
Under the modular symmetry of the modulus $u$, 
the behavior of fluxes is different by factors from those of $T^6/(\mathbb{Z}_2\times\mathbb{Z}_2')$ in the previous section.
The $S$ transformation corresponds to the following replacement of fluxes:
\begin{align}
      e_0&\rightarrow-\frac{m_0}{6}, & e&\rightarrow \frac{m}{2}, & m&\rightarrow-2e, & m_0&\rightarrow 6e_0, 
\end{align}
and $T$ transformation corresponds to the shift of fluxes,
\begin{align}
    e_0&\rightarrow e_0+e+\frac{m}{2}+\frac{m_0}{6}, & e&\rightarrow e+m+\frac{m_0}{2}, & m&\rightarrow m+m_0, & m_0&\rightarrow m_0.
\end{align}

{These transformations are not consistent with flux quantization for all fluxes.
To be consistent, the modular symmetry is constrained to a subgroup of $SL(2,\mathbb{Z})$.\footnote{See for remaining modular symmetry due to flux background,  e.g. Ref.~\cite{Kobayashi:2020hoc}.}}
{In the normalization of Eq.~(\ref{eq:cicy_def}), the integer flux vector is $(e_0,e,m,m_0)\in\mathbb{Z}^4$, while the superpotential is expanded in the basis}
\begin{align}
{W(u)=e_0+eu+\frac{m}{2}u^2+\frac{m_0}{6}u^3.}
\end{align}
{For a general modular transformation}
\begin{align}
{\gamma=\begin{pmatrix}a&b\\c&d\end{pmatrix}\in SL(2,\mathbb{Z}),}
\end{align}
{integrality of the transformed coefficients for every integer flux vector requires $b=0\pmod 6$. Thus, in this normalization, the subgroup preserving the physical integral flux lattice is}
\begin{align}
{\Gamma^0(6)=\left\{\left.\begin{pmatrix}a&b\\c&d\end{pmatrix}\in SL(2,\mathbb{Z})\ \right|\ b=0\pmod 6\right\}.}
\end{align}
{Thus, the transformations displayed above express the formal modular covariance of the polynomial, while only $\Gamma^0(6)$ acts within the quantized flux ensemble.}

\subsection{Distribution of modulus vacua}

The solutions of Eqs.~(\ref{eq:cicy_alpha}) and (\ref{eq:cicy_beta}) for $-N \leq e_0,e,m,m_0 \leq N$ become specific sequence with $\alpha$ fixed.
Table \ref{tab:beta-CICY-0} shows $\beta^2$ with $\alpha=0$ when we vary $N$ from 1 to 10.
They are large compared with those in $T^6/(\mathbb{Z}_2\times\mathbb{Z}_2')$.

\begin{longtable}{ll}
    \caption{$\beta^2$ sequence with $\alpha=0$ for \(-N \leq e_0, e, m, m_0 \leq N\). }
    \label{tab:beta-CICY-0}\\ \hline
    $N$ & $\beta^2~(\alpha=0)$ \\ \hline
    \endfirsthead
    \hline
    $N$ & $\beta^2~(\alpha=0)$ \\ \hline
    \endhead
    \hline
    \endfoot
    \hline
    \endlastfoot
\(1\) & \(\frac{2}{1}\), \(\frac{6}{1}\) \\
\(2\) & \(\frac{1}{1}\), \(\frac{2}{1}\), \(\frac{3}{1}\),
\(\frac{4}{1}\), \(\frac{6}{1}\), \(\frac{12}{1}\) \\
\(3\) & \(\frac{2}{3}\), \(\frac{1}{1}\), \(\frac{4}{3}\),
\(\frac{2}{1}\), \(\frac{3}{1}\), \(\frac{4}{1}\), \(\frac{6}{1}\),
\(\frac{9}{1}\), \(\frac{12}{1}\), \(\frac{18}{1}\) \\
\(4\) & \(\frac{1}{2}\), \(\frac{2}{3}\), \(\frac{1}{1}\),
\(\frac{4}{3}\), \(\frac{3}{2}\), \(\frac{2}{1}\), \(\frac{8}{3}\),
\(\frac{3}{1}\), \(\frac{4}{1}\), \(\frac{9}{2}\), \(\frac{6}{1}\),
\(\frac{8}{1}\), \(\frac{9}{1}\), \(\frac{12}{1}\), \(\frac{18}{1}\),
\(\frac{24}{1}\) \\
\(5\) & \(\frac{2}{5}\), \(\frac{1}{2}\), \(\frac{2}{3}\),
\(\frac{4}{5}\), \(\frac{1}{1}\), \(\frac{6}{5}\), \(\frac{4}{3}\),
\(\frac{3}{2}\), \(\frac{8}{5}\), \(\frac{2}{1}\), \(\frac{12}{5}\),
\(\frac{5}{2}\), \(\frac{8}{3}\), \(\frac{3}{1}\), \(\frac{10}{3}\),
\(\frac{18}{5}\), \(\frac{4}{1}\), \(\frac{9}{2}\), \(\frac{24}{5}\),
\(\frac{5}{1}\), \(\frac{6}{1}\), \(\frac{15}{2}\), \(\frac{8}{1}\),
\(\frac{9}{1}\), \(\frac{10}{1}\), \(\frac{12}{1}\), \(\frac{15}{1}\), \\ &
\(\frac{18}{1}\), \(\frac{24}{1}\), \(\frac{30}{1}\) \\
\(6\) & \(\frac{1}{3}\), \(\frac{2}{5}\), \(\frac{1}{2}\),
\(\frac{2}{3}\), \(\frac{4}{5}\), \(\frac{1}{1}\), \(\frac{6}{5}\),
\(\frac{4}{3}\), \(\frac{3}{2}\), \(\frac{8}{5}\), \(\frac{5}{3}\),
\(\frac{2}{1}\), \(\frac{12}{5}\), \(\frac{5}{2}\), \(\frac{8}{3}\),
\(\frac{3}{1}\), \(\frac{10}{3}\), \(\frac{18}{5}\), \(\frac{4}{1}\),
\(\frac{9}{2}\), \(\frac{24}{5}\), \(\frac{5}{1}\), \(\frac{6}{1}\),
\(\frac{36}{5}\), \(\frac{15}{2}\), \(\frac{8}{1}\), \(\frac{9}{1}\),
\(\frac{10}{1}\), \\ & \(\frac{12}{1}\), \(\frac{15}{1}\), \(\frac{18}{1}\),
\(\frac{24}{1}\), \(\frac{30}{1}\), \(\frac{36}{1}\) \\
\(7\) & \(\frac{2}{7}\), \(\frac{1}{3}\), \(\frac{2}{5}\),
\(\frac{1}{2}\), \(\frac{4}{7}\), \(\frac{2}{3}\), \(\frac{4}{5}\),
\(\frac{6}{7}\), \(\frac{1}{1}\), \(\frac{8}{7}\), \(\frac{6}{5}\),
\(\frac{4}{3}\), \(\frac{10}{7}\), \(\frac{3}{2}\), \(\frac{8}{5}\),
\(\frac{5}{3}\), \(\frac{12}{7}\), \(\frac{2}{1}\), \(\frac{7}{3}\),
\(\frac{12}{5}\), \(\frac{5}{2}\), \(\frac{18}{7}\), \(\frac{8}{3}\),
\(\frac{14}{5}\), \(\frac{3}{1}\), \(\frac{10}{3}\), \(\frac{24}{7}\),
\(\frac{7}{2}\), \\ &  \(\frac{18}{5}\), \(\frac{4}{1}\), \(\frac{30}{7}\),
\(\frac{9}{2}\), \(\frac{14}{3}\), \(\frac{24}{5}\), \(\frac{5}{1}\),
\(\frac{36}{7}\), \(\frac{6}{1}\), \(\frac{7}{1}\), \(\frac{36}{5}\),
\(\frac{15}{2}\), \(\frac{8}{1}\), \(\frac{42}{5}\), \(\frac{9}{1}\),
\(\frac{10}{1}\),  \(\frac{21}{2}\), \(\frac{12}{1}\), \(\frac{14}{1}\),
\(\frac{15}{1}\), \(\frac{18}{1}\), \(\frac{21}{1}\), \(\frac{24}{1}\),
\(\frac{30}{1}\), \\ & \(\frac{36}{1}\), \(\frac{42}{1}\) \\
\(8\) & \(\frac{1}{4}\), \(\frac{2}{7}\), \(\frac{1}{3}\),
\(\frac{2}{5}\), \(\frac{1}{2}\), \(\frac{4}{7}\), \(\frac{2}{3}\),
\(\frac{3}{4}\), \(\frac{4}{5}\), \(\frac{6}{7}\), \(\frac{1}{1}\),
\(\frac{8}{7}\), \(\frac{6}{5}\), \(\frac{5}{4}\), \(\frac{4}{3}\),
\(\frac{10}{7}\), \(\frac{3}{2}\), \(\frac{8}{5}\), \(\frac{5}{3}\),
\(\frac{12}{7}\), \(\frac{7}{4}\), \(\frac{2}{1}\), \(\frac{9}{4}\),
\(\frac{16}{7}\), \(\frac{7}{3}\), \(\frac{12}{5}\), \(\frac{5}{2}\),
\(\frac{18}{7}\), \(\frac{8}{3}\), \\ & \(\frac{14}{5}\), \(\frac{3}{1}\),
\(\frac{16}{5}\), \(\frac{10}{3}\), \(\frac{24}{7}\), \(\frac{7}{2}\),
\(\frac{18}{5}\), \(\frac{15}{4}\), \(\frac{4}{1}\), \(\frac{30}{7}\),
\(\frac{9}{2}\), \(\frac{14}{3}\), \(\frac{24}{5}\), \(\frac{5}{1}\),
\(\frac{36}{7}\), \(\frac{21}{4}\), \(\frac{16}{3}\), \(\frac{6}{1}\),
\(\frac{48}{7}\), \(\frac{7}{1}\), \(\frac{36}{5}\), \(\frac{15}{2}\),
\(\frac{8}{1}\), \(\frac{42}{5}\), \(\frac{9}{1}\), \\ & \(\frac{48}{5}\),
\(\frac{10}{1}\), \(\frac{21}{2}\), \(\frac{12}{1}\), \(\frac{14}{1}\),
\(\frac{15}{1}\), \(\frac{16}{1}\), \(\frac{18}{1}\), \(\frac{21}{1}\),
\(\frac{24}{1}\), \(\frac{30}{1}\), \(\frac{36}{1}\), \(\frac{42}{1}\),
\(\frac{48}{1}\) \\
\(9\) & \(\frac{2}{9}\), \(\frac{1}{4}\), \(\frac{2}{7}\),
\(\frac{1}{3}\), \(\frac{2}{5}\), \(\frac{4}{9}\), \(\frac{1}{2}\),
\(\frac{4}{7}\), \(\frac{2}{3}\), \(\frac{3}{4}\), \(\frac{4}{5}\),
\(\frac{6}{7}\), \(\frac{8}{9}\), \(\frac{1}{1}\), \(\frac{10}{9}\),
\(\frac{8}{7}\), \(\frac{6}{5}\), \(\frac{5}{4}\), \(\frac{4}{3}\),
\(\frac{10}{7}\), \(\frac{3}{2}\), \(\frac{14}{9}\), \(\frac{8}{5}\),
\(\frac{5}{3}\), \(\frac{12}{7}\), \(\frac{7}{4}\), \(\frac{16}{9}\),
\(\frac{2}{1}\), \(\frac{9}{4}\), \\ & \(\frac{16}{7}\), \(\frac{7}{3}\),
\(\frac{12}{5}\), \(\frac{5}{2}\), \(\frac{18}{7}\), \(\frac{8}{3}\),
\(\frac{14}{5}\), \(\frac{3}{1}\), \(\frac{16}{5}\), \(\frac{10}{3}\),
\(\frac{24}{7}\), \(\frac{7}{2}\), \(\frac{18}{5}\), \(\frac{15}{4}\),
\(\frac{4}{1}\), \(\frac{30}{7}\), \(\frac{9}{2}\), \(\frac{14}{3}\),
\(\frac{24}{5}\), \(\frac{5}{1}\), \(\frac{36}{7}\), \(\frac{21}{4}\),
\(\frac{16}{3}\), \(\frac{6}{1}\), \(\frac{27}{4}\), \\ &  \(\frac{48}{7}\),
\(\frac{7}{1}\), \(\frac{36}{5}\), \(\frac{15}{2}\), \(\frac{54}{7}\),
\(\frac{8}{1}\), \(\frac{42}{5}\), \(\frac{9}{1}\), \(\frac{48}{5}\),
\(\frac{10}{1}\), \(\frac{21}{2}\), \(\frac{54}{5}\), \(\frac{12}{1}\),
\(\frac{27}{2}\), \(\frac{14}{1}\), \(\frac{15}{1}\), \(\frac{16}{1}\),
\(\frac{18}{1}\), \(\frac{21}{1}\), \(\frac{24}{1}\), \(\frac{27}{1}\),
\(\frac{30}{1}\), \(\frac{36}{1}\), \(\frac{42}{1}\), \\ &  \(\frac{48}{1}\),
\(\frac{54}{1}\) \\
\(10\) & \(\frac{1}{5}\), \(\frac{2}{9}\), \(\frac{1}{4}\),
\(\frac{2}{7}\), \(\frac{1}{3}\), \(\frac{2}{5}\), \(\frac{4}{9}\),
\(\frac{1}{2}\), \(\frac{4}{7}\), \(\frac{3}{5}\), \(\frac{2}{3}\),
\(\frac{3}{4}\), \(\frac{4}{5}\), \(\frac{6}{7}\), \(\frac{8}{9}\),
\(\frac{1}{1}\), \(\frac{10}{9}\), \(\frac{8}{7}\), \(\frac{6}{5}\),
\(\frac{5}{4}\), \(\frac{4}{3}\), \(\frac{7}{5}\), \(\frac{10}{7}\),
\(\frac{3}{2}\), \(\frac{14}{9}\), \(\frac{8}{5}\), \(\frac{5}{3}\),
\(\frac{12}{7}\), \(\frac{7}{4}\), \\ & \(\frac{16}{9}\), \(\frac{9}{5}\),
\(\frac{2}{1}\), \(\frac{20}{9}\), \(\frac{9}{4}\), \(\frac{16}{7}\),
\(\frac{7}{3}\), \(\frac{12}{5}\), \(\frac{5}{2}\), \(\frac{18}{7}\),
\(\frac{8}{3}\), \(\frac{14}{5}\), \(\frac{20}{7}\), \(\frac{3}{1}\),
\(\frac{16}{5}\), \(\frac{10}{3}\), \(\frac{24}{7}\), \(\frac{7}{2}\),
\(\frac{18}{5}\), \(\frac{15}{4}\), \(\frac{4}{1}\), \(\frac{21}{5}\),
\(\frac{30}{7}\), \(\frac{9}{2}\), \(\frac{14}{3}\), \(\frac{24}{5}\), \\ & 
\(\frac{5}{1}\), \(\frac{36}{7}\), \(\frac{21}{4}\), \(\frac{16}{3}\),
\(\frac{27}{5}\), \(\frac{6}{1}\), \(\frac{20}{3}\), \(\frac{27}{4}\),
\(\frac{48}{7}\), \(\frac{7}{1}\), \(\frac{36}{5}\), \(\frac{15}{2}\),
\(\frac{54}{7}\), \(\frac{8}{1}\), \(\frac{42}{5}\), \(\frac{60}{7}\),
\(\frac{9}{1}\), \(\frac{48}{5}\), \(\frac{10}{1}\), \(\frac{21}{2}\),
\(\frac{54}{5}\), \(\frac{12}{1}\), \(\frac{27}{2}\), \(\frac{14}{1}\),
\(\frac{15}{1}\), \\ & \(\frac{16}{1}\), \(\frac{18}{1}\), \(\frac{20}{1}\),
\(\frac{21}{1}\), \(\frac{24}{1}\), \(\frac{27}{1}\), \(\frac{30}{1}\),
\(\frac{36}{1}\), \(\frac{42}{1}\), \(\frac{48}{1}\), \(\frac{54}{1}\),
\(\frac{60}{1}\) \\
\end{longtable}

Similar to the $T^6/(\mathbb{Z}_2\times\mathbb{Z}_2')$, we can understand these sequences.
We define
\begin{align}
    M(k)=
    \begin{pmatrix}
        6e_0 & m \\
        2e  & m_0
    \end{pmatrix}.
\end{align}
When ${\rm Re}~u=0$, we find $B(k)={\det} M(k)=0$.
Hence, we can write 
\begin{align}
    \begin{pmatrix}
        6e_0 \\ 2e
    \end{pmatrix}
    =\lambda 
    \begin{pmatrix}
        r \\ s
    \end{pmatrix}, \qquad 
    \begin{pmatrix}
        m \\ m_0
    \end{pmatrix}
    =\mu
    \begin{pmatrix}
        r \\ s
    \end{pmatrix}.
\end{align}
Then, we obtain the same equation as Eq.~(\ref{eq:lambda/mu}), $\beta^2=-\lambda/\mu$.
The maximum value of $\beta^2=-\lambda/\mu$ for $-N \leq e_0,e,m,m_0 \leq N$ is obtained when $s=0$, $e_0=\pm N$, and $m=\mp1$ (1), i.e., $\lambda=\pm 6N$ and $\mu=\mp1$.
That corresponds to $\beta^2=6N$.
On the other hand, the minimum value of $\beta^2=-\lambda/\mu$ is obtained when $r=0$, $e=\pm 1$ and $m_0=\mp N$, i.e., $\lambda=\pm 1$ and $\mu = \mp N$.
That corresponds to {$\beta^2=2/N$}.
Similarly, other values between {$2/N$} and $6N$ in the sequence for $N$ are obtained.
Indeed, by multiplying  2 or 6 by the numbers in 
the sequences of Table~\ref{tab:beta-orientifold-0}, we can obtain the corresponding numbers of $\beta^2$ in 
Table~\ref{tab:beta-CICY-0}.
For example, for $N=1$ in Table~\ref{tab:beta-orientifold-0}, we have only $\beta^2=1$.
We multiply 2 and 6 by $\beta^2=1$.
Then, we obtain $\beta^2=2$ and 6, respectively, which are $\beta^2$ for $N=1$ in Table~\ref{tab:beta-CICY-0}.
For $N=2$, we have $1/2$, 1, and 2 in Table~\ref{tab:beta-orientifold-0}.
When we multiply them by 2, we obtain
\begin{align}
    1, 2, 4.
\end{align}
When we multiply them by 6, we obtain 
\begin{align}
    3, 6, 12.
\end{align}
The set including these numbers correspond to $\beta^2$ for $N=2$ in Table~\ref{tab:beta-CICY-0}.
Moreover, for $N=3$, we have 
\begin{align}
    \frac13, \frac12, \frac23,1,\frac32,2,3,
\end{align}
in Table~\ref{tab:beta-orientifold-0}.
When we multiply them by 2, we obtain
\begin{align}
    \frac23, 1,\frac43,2,3,4,6.
\end{align}
When we multiply them by 6, we obtain
\begin{align}
    2,3,4,6,9,12,18.
\end{align}
Their union corresponds to $\beta^2$ for $N=3$ in Table~\ref{tab:beta-CICY-0}.
In general, the numbers of $\beta^2\leq 1 $ for $N$ in Table~\ref{tab:beta-orientifold-0} correspond to the Farey sequence $F_N$.
The numbers of $\beta^2\geq1$ for $N$ in the table are their inverses.
Let us denote them by $F^{-1}_N$.
The above observation implies that when we multiply 2 and 6 by all the numbers in $F_N \cup F^{-1}_N$, we obtain $\beta^2$ for $N$ in Table~\ref{tab:beta-CICY-0}.
In this sense, the sequence in Table~\ref{tab:beta-CICY-0} is related to the Farey sequence.
It is remarkable that a single value $\beta^2$ in Table~\ref{tab:beta-orientifold-0} corresponds to one or two values in Table~\ref{tab:beta-CICY-0}.
In this sense, degeneracy is resolved.
Indeed, various values of $u$ are realized in this Calabi-Yau compactification compared with $T^6/(\mathbb{Z}_2\times\mathbb{Z}_2')$.
Values of $\beta^2$ become larger.

\subsection{CP violation}

As in $T^6/(\mathbb{Z}_2\times\mathbb{Z}_2')$, we can discuss the possibility for spontaneous CP violation.
There are two possibilities for the CP symmetric $|W|^2$.
The first one is the case with $e_0=m=0$, and the second one is the case with $e=m_0=0$.
We require ${\rm Im}~u>0$ as a physical condition.
Then, both cases lead to the solution ${\rm Re}~u=0$.
The vacuum is always CP symmetric.

\section{Multi-moduli models and CP violation}
\label{sec:multi-moduli}

We can extend the previous analysis on a single modulus to the case with {multiple} moduli, \(u^a = \alpha^a + i\beta^a\).
Their K\"ahler potential is written by 
\begin{equation}
K = - \ln \kappa, \quad \kappa = \frac{i}6 \kappa_{abc} (u^a-\bar u^a) (u^b-\bar u^b) (u^c-\bar u^c) = \frac43\kappa_{abc} \beta^a \beta^b \beta^c,
\end{equation}
and the superpotential is written by 
\begin{equation}
W(u^a) = e_0 + e_a u^a + \frac12 m^c \kappa_{abc} u^a u^b + \frac16 m^0 \kappa_{abc} u^a u^b u^c.
\end{equation}

The first derivative of the superpotential is obtained by 
\begin{equation}
W_a = e_a + m^c \kappa_{abc} u^b + \frac12 m^0 \kappa_{abc} u^b u^c,
\end{equation}
which is decomposed to the real part and imaginary part as follows,
\begin{equation}\label{eq:ReWi_ImWi}
\begin{cases}
  {\rm Re} W_a = e_a + m^c \kappa_{abc} \alpha^b + \frac12 m^0 \kappa_{abc} (\alpha^b \alpha^c - \beta^b \beta^c), \\
  {\rm Im} W_a = m^c \kappa_{abc} \beta^b + m^0 \kappa_{abc} \alpha^b \beta^c.
\end{cases}
\end{equation}
The second derivative is also written by
\begin{equation}
W_{ab} = m^c \kappa_{abc} + m^0 \kappa_{abc} u^c.
\end{equation}

\subsection{SUSY minimum}
\label{sec:multi-SUSY}

The SUSY condition is written by 
\begin{equation}\label{eq:DiW}
D_a W = W_a + K_a W = W_a - \frac{\kappa_a}{\kappa} W = 0,
\end{equation}
which can be written as follows,
\begin{equation}
W_a = \frac{W}{\kappa} \kappa_a = c \kappa_{abc} \beta^b \beta^c, \quad c \equiv - \frac{2iW}{\kappa}.
\end{equation}

When we require $\kappa \neq 0$, the above equation becomes 
\begin{equation}\label{eq:omega_wi_omwgai_w}
\kappa W_a - \kappa_a W = 0.
\end{equation} 
This equation is decomposed to the real part and imaginary part as follows,
\begin{equation}\label{eq:ReWi_ImWi_cond}
\begin{cases}
  {\rm Re} W_a = ({\rm Re} ~c) \kappa_{abc} \beta^b \beta^c, \\
  {\rm Im} W_a = ({\rm Im} ~c) \kappa_{abc} \beta^b \beta^c.
\end{cases}
\end{equation}
We multiply $(u^a-\bar u^a)$ by both sides in Eq.~(\ref{eq:omega_wi_omwgai_w}) so as to derive 
\begin{align}
    \kappa(u^a-\bar u^a)  W_a - \kappa_a (u^a-\bar u^a)W = 0.
\end{align}
By using the identity \((u^a-\bar u^a) \kappa_a = 3\kappa\), we obtain 
\begin{align}\label{eq:multi-SUSY}
    (u^a-\bar u^a) W_a - 3 W = 0.
\end{align}

Now we introduce a new field $u^0=t$
such that we write the superpotential by the following homogeneous one:
\begin{equation}
\begin{aligned}
  \mathcal{W}(u^i)
    &= \frac16 \mathcal{W}_{ijk} u^i u^j u^k \\
    &= e_0 t^3 + e_a t^2 u^a + \frac12 m^c \kappa_{abc} t u^a u^b + \frac16 m^0 \kappa_{abc} u^a u^b u^c.
\end{aligned}
\end{equation}
We denote \(u^i = (t, u^a)\).
Note that $\mathcal{W}(t=1,u^a)=W(u^a)$.
Similar to Eq.~(\ref{eq:multi-SUSY}), the SUSY condition can be written by 
\begin{equation}
(u^i-\bar u^i) \mathcal{W}_i - 3 \mathcal{W} = 0.
\end{equation}
We use the identity $u^i\mathcal{W}_i=3\mathcal{W}$.
Then the above condition can be written by
\begin{equation}
\bar u^i \mathcal{W}_i = 0.
\end{equation}
Furthermore, by use of the identities \(\mathcal{W}_{ij} u^j = 2\mathcal{W}_i\),
\(\mathcal{W}_{ijk} u^k = \mathcal{W}_{ij}\), the condition can be written by 
 \begin{equation}
\mathcal{W}_{ijk} u^i u^j {\bar u}^k = 0, \quad \mathcal{W}_{ijk} u^i {\bar u}^j {\bar u}^k = 0.
\end{equation}

Here, we define 
\begin{equation}\label{eq:w_a}
w_i = \mathcal{W}_{ijk} u^j {\bar u}^k = \mathcal{W}_{ij} {\bar u}^j.
\end{equation}
By using this vector $w_i$, we can write the condition as follows,
\begin{equation}
w_i u^i = 0, \quad w_i {\bar u}^i = 0,
\end{equation}
which can be decomposed to the real part and imaginary part,
\begin{equation}\label{eq:w_alpha_w_beta}
w_i \alpha^i = 0, \quad w_i \beta^i = 0.
\end{equation} 

Here we set $u^0=t=1$.
The condition (\ref{eq:w_alpha_w_beta}) is written by 
\begin{align}
    w_0 + w_a \alpha^a = 0, \quad w_a \beta^a = 0.
\end{align}
By use of $\bar u^i=u^i-2i\beta^i$, the vector $w_a$ can be written explicitly by
\begin{align}
    \label{eq:wa_rewrite}
w_a &= \mathcal{W}_{ab} u^b - 2i \mathcal{W}_{ab} \beta^b \notag \\ &= 2\mathcal{W}_a - 2i \mathcal{W}_{ab} \beta^b \notag \\
&= 2 W_a - 2i(m^c \kappa_{abc} \beta^b + m^0 \kappa_{abc} \alpha^c \beta^b) + 2 m^0 \kappa_{abc} \beta^b \beta^c \notag \\
&= 2 {\rm Re} W_a + 2 m^0 \kappa_{abc} \beta^b \beta^c. 
\end{align}
Using Eq.~(\ref{eq:ReWi_ImWi}), we can write the condition as 
\begin{align}
    w_a = 2 ({\rm Re}~ c + m^0) \kappa_{abc} \beta^b \beta^c.
\end{align}
We multiply both sides of the above equation by $\beta^a$ so as to find 
\begin{align}
    w_a\beta^a = 2 ({\rm Re}~ c + m^0) \kappa_{abc} \beta^a\beta^b \beta^c=\frac32 ({\rm Re}~c+m^0)\kappa.
\end{align}
Note that $w_a \beta^a=0$ because of the SUSY condition and we require $\kappa >0$ as a physical condition.
Then, we find $({\rm Re}~c+m^0)=0$.
Finally, we obtain the condition,
\begin{align}\label{eq:SUSY-final}
    w_0 = 0, \quad w_a = 0.
\end{align}
By use of these conditions, we can study vacua of multi moduli models.
A detailed analysis on multi modulus like one in Section \ref{sec:orientifol} is beyond our scope.
Instead, here we study the possibility for spontaneous CP violation.

\subsection{Spontaneous CP violation}
\label{sec:multi-CP}

Here we study the possibility for spontaneous CP violation by use of the condition (\ref{eq:SUSY-final}).
There are two possibilities for the CP symmetric $|W|^2$.
The first one is $m_0=e=0$, while the second one is $m=e_0=0$.

For the first one, the superpotential and its derivatives are  written by 
\begin{equation}
W(u^a) = e_0 + \frac12 M_{ab} u^a u^b, \quad M_{ab} \equiv m^c \kappa_{abc}, \quad W_a = M_{ab} u^b, \quad W_{ab} = M_{ab}.
\end{equation}
Then, the vector $w_i$ are obtained explicitly by
\begin{equation}
w_0 = 6e_0 + M_{ab} (\alpha^a\alpha^b + \beta^a\beta^b), \quad
w_a = 2M_{ab} \alpha^b.
\end{equation}
The SUSY condition (\ref{eq:SUSY-final}) requires $w_0=w_a=0$.
If \(\det M_{ab} = \det (m^c \kappa_{abc}) \neq 0\), $\alpha^a$ must vanish, $\alpha^a=0$.
This is a {CP-symmetric vacuum}.
If $\det M=0$, there are flat directions among $\alpha^a$ along vanishing eigenvalues of $M_{ab}$.
Along such flat directions, the CP symmetric vacuum and CP violating vacuum are degenerate.

For the second case with $e_0=m^a=0$, 
the superpotential is written by 
\begin{equation}
W(u^a) = e_a u^a + \frac16 m^0 \kappa_{abc} u^a u^b u^c.
\end{equation}
The vector $w_i$ is obtained explicitly as 
\begin{equation}
w_0 = 4e_a \alpha^a, \quad w_a = 2e_a + m^0 \kappa_{abc} (\alpha^b\alpha^c + \beta^b\beta^c).
\end{equation}
Then, the SUSY condition $(\ref{eq:SUSY-min})$ requires $w_0=w_a=0$, i.e.,
\begin{align}
&  e_a \alpha^a = 0, \notag\\
&  2e_a + m^0 \kappa_{abc} (\alpha^b\alpha^c + \beta^b\beta^c) = 0.
\end{align}
By multiplying the second equation by $\alpha^a$, we obtain
\begin{equation}\label{eq:alpha_beta_beta}
m^0 \kappa_{abc} \alpha^a\alpha^b\alpha^c + m^0 \kappa_{abc} \alpha^a\beta^b\beta^c = 0.
\end{equation}
On the other hand, we can write ${\rm Im}~W_a$ as  \({\rm Im} W_a = m^0 \kappa_{abc} \alpha^b \beta^c\) from Eq.~(\ref{eq:ReWi_ImWi}) and \({\rm Im} W_a = ({\rm Im} ~c) \kappa_{abc} \beta^b \beta^c\) from Eq.~(~\ref{eq:ReWi_ImWi_cond}).
Hence, we find 
\begin{equation}
\kappa_{abc} \beta^c [m^0 \alpha^b - ({\rm Im}~ c) \beta^b] = 0.
\end{equation}
$m_0$ must not vanish.
Otherwise, the superpotential becomes $W=e_au^a$, which is not suitable because the SUSY condition requires $e_a=0$.\footnote{Also, the linear superpotential terms such as $W=e_au^a$ without mass terms can not stabilize modulus \cite{Abe:2006xi}.}
If $\det (\kappa_{abc} \beta^c )\neq 0$, we find 
\begin{equation}
\alpha^a = \rho \beta^a, \quad \rho \equiv \frac{{\rm Im}~ c}{m^0} \in \mathbb{R}.
\end{equation}
We substitute this relation to Eq.~(\ref{eq:alpha_beta_beta}) so as to derive the following equation:
\begin{equation}
m^0 \rho(1+\rho^2) \kappa_{abc} \beta^a\beta^b\beta^c = \frac34 m^0 \rho(1+\rho^2) \kappa = 0.
\end{equation}
As a result, we find $\alpha^a=0$.
That is, the vacuum is CP-symmetric.

To derive the above result, we assume 
$\det (\kappa_{abc} \beta^c )\neq 0$.
Let us examine the meaning of this assumption.
Suppose that \(\kappa_{abc} \beta^c v^b = 0\) for some vector $v^b$.
This vector also satisfies \(K_{a\bar b} v^b = 0\).
Then, it is found that 
\begin{equation}
\ker (\kappa_{abc}\beta^c) \subseteq \ker K_{a\bar b}.
\end{equation}
That implies that if \(\det K_{a\bar b} \neq 0\), we always have 
\(\det (\kappa_{abc} \beta^c) \neq 0\).
The vacuum is always CP symmetric for proper K\"ahler metric with \(\det K_{a\bar b} \neq 0\).

\section{Conclusions}
\label{sec:conclusion}

We have studied modulus stabilization by background fluxes. 
{In the truncated complex-structure sector, the F-term condition reduces to the holomorphic quadratic equation $F(u)=0$.} 
As concrete examples, we have considered the $T^6/(\mathbb{Z}_2\times\mathbb{Z}_2')$ orientifold and a simple complete intersection Calabi-Yau compactification.
The modulus vacua show specific patterns.
For fixed ${\rm Re}~u$, $({\rm Im}~u)^2$ exhibits specific sequences.
We find the Farey sequence for $({\rm Im}~u)^2$ in $T^6/(\mathbb{Z}_2\times\mathbb{Z}_2')$ with ${\rm Re}~u=0$. 
In the Calabi-Yau compactification, we can find a similar sequence, which is related to the toroidal Farey data through the numerical factors appearing in the flux-induced superpotential. 
For other values of ${\rm Re}~u$, we find specific sequences.

A void structure similar to that found in Ref.~\cite{Denef:2004ze} also appears in the distribution of the modulus vacua. Highly degenerate vacua are located at the centers of the voids. We find a numerical correlation between their degeneracies and the squared distance to the nearest neighboring vacuum, which is used as a measure of the void size. 
{We find a similar correlation for different values of $N$, although its slope changes with $N$. The detailed void structure also depends on the flux cutoff and on how the fluxes are sampled.}

For a fixed negative discriminant and primitive coefficients, the corresponding equivalence classes are related by Gauss composition and form a finite Abelian class group, such as $\mathbb{Z}_N \times \mathbb{Z}_{N'} \times \cdots$. The CP transformation acts on this class group by inversion,
\begin{align*}
[A,B,C]\longmapsto[A,-B,C]=[A,B,C]^{-1}.
\end{align*}
Thus, Gauss composition provides an arithmetic organization of the modulus vacua and their CP conjugates. Whether this class-group action can be promoted to a physical symmetry of a single four-dimensional effective theory requires further investigation.

We have also studied CP violation. 
{In the one-modulus examples, a CP-symmetric flux superpotential leads to a CP-symmetric solution of the F-term condition for the complex structure moduli, and an isolated solution with spontaneous CP violation is not obtained.} In multi-modulus models, the same conclusion follows when the relevant flux and intersection matrices are nondegenerate. When these matrices are degenerate, flat directions may remain, along which CP-preserving and CP-violating configurations can be degenerate. 
{
For the RR-flux superpotential, the one-modulus solutions lie in imaginary quadratic number fields. With NS fluxes and the axio-dilaton, however, the F-term equations for $u$ and $s$ are coupled, and the resulting equation for $u$ need not remain quadratic. It remains to be seen whether imaginary quadratic fields still appear on special loci or are replaced by higher-degree number fields. The extension to several complex structure moduli, together with tadpole constraints, quantum corrections, and non-supersymmetric extrema, is left for future work.
}

\acknowledgments

This work was supported by JSPS KAKENHI Grant Numbers JP23K03375 (T.K.), JP25H01539 (H.O.) and JP26K07087 (H.O.), and JST SPRING, Grant Number JPMJSP2119 (R.N.).

\appendix

\section{Modular forms}
\label{app:modular-form}

The holomorphic quadratic equation $F(u)=0$ plays an important role in discussions of modulus vacua.
Here, we discuss another role of the holomorphic quadratic equation in the modular symmetry.

We consider the equivalence class $[A_0,B_0,C_0]$, which satisfies 
\begin{gather}
    A_0 > 0,\quad \rm{gcd}(A_0,B_0,C_0)=1, \notag \\
    \quad D_0 \equiv B_0^2 - 4A_0C_0 < 0.
\end{gather}
We write the holomorphic function with these coefficients,
\begin{align}
    F_0(X,Y)\equiv A_0X^2+B_0XY+C_0Y^2.
\end{align}
Here, we define the modular transformation of $F_0(X,Y)$ by 
\begin{align}
    \gamma : F_0(X,Y)\to F_0(aX+bY,cX+dY)\equiv F_0(X,Y)|_{\gamma},
\end{align}
where 
\begin{align}
    \gamma = \begin{pmatrix}
    a & b\\
    c & d
\end{pmatrix}
\in SL(2,\mathbb{Z}).
\end{align}
Note that 
\begin{align}
    u_0 \equiv \frac{-B_0+i\sqrt{-D_0}}{2A_0}
\end{align}
is a solution of $F_0(u_0,1)=0$.

Now, let us define the following holomorphic function:
\begin{equation}
    f_{s,[F_0]}(u) =
    \zeta(s)
    \sum_{\gamma\in SL(2,\mathbb{Z})/F_{inv}} \frac1{(F_0(u,1)|_\gamma)^s},
\end{equation}
where $F_{inv}$ denotes the modular transformations, under which  $F_0(X,Y)$ is invariant, i.e., 
\begin{align}
   F_{inv} \equiv \{ \gamma\in SL(2,\mathbb{Z})\ \ {\rm{s.t.}}\ F_0(X,Y)|_{\gamma} = F_0(X,Y) \} .
\end{align}
The holomorphic function $f_{s,[F_0]}(u)$ satisfies
\begin{equation}
    f_{s,[F_0]}(\gamma\tau) = (c\tau+d)^{2s} f_{s,[F_0]}(\tau),
\end{equation}
under the modular symmetry.
It behaves as a meromorphic modular form of the weight $2s$.
(See e.g. \cite{Bengoechea}.)
It has a pole at $u=u_0$ in the upper half plane, while $f_{s,[F_0]} \to 0$ at the limit ${\rm Im}~u \to \infty$.
For example, when $[A_0,B_0,C_0]=[1,1,1]$, $u_0$ corresponds to $e^{2\pi i /3}$, and the above function has a pole at this point.
For instance, the function of weight 4 can be written by \cite{Bengoechea},
\begin{align}
    f_{2,[1,1,1]}=\frac{\Delta(u)}{E_4(u)^2},
\end{align}
up to normalization, where $\Delta(u)=(E_4^3(u)-E_6(u)^2)/1728$ and $E_k(u)$ denotes Eisenstein series of weight $k$.  
Similarly, $[A_0,B_0,C_0]=[1,0,1]$, $u_0$ corresponds to $u_0=i$, and the above function has a pole at this point.
For example, the function of weight 4 \cite{Bengoechea}
\begin{align}
    f_{2,[1,0,1]}=\frac{\Delta(u)E_4(u)}{E_6(u)^2},
\end{align}
up to normalization.
Other functions with poles at different points from $u=i$ and $e^{2\pi i/3}$, can also be written by use of Eisenstein series and $j$-function.

Now, let us assume the following superpotential:
\begin{equation}
    \widetilde{W}_{s,[F_0]} = \frac1{f_{s,[F_0]}}.
\end{equation}
It transforms as 
\begin{equation}
    \widetilde{W}_{s,[F_0]}(\gamma u) = (cu+d)^{-2s} \widetilde{W}_{s,[F_0]}(u),
\end{equation}
under the modular symmetry.
Because it satisfies 
\begin{equation}
    \widetilde{W}_{s,[F_0]}(u_Q) = 0, \quad \partial_{u} \widetilde{W}_{s,[F_0]}(u_Q) = 0,
\end{equation}
it satisfies the SUSY condition $D_uW=0$ at $u=u_Q$.
Thus, the superpotential $\widetilde{W}_{s,[F_0]}$ leads to the same minimum as the flux-induced superpotential with the flux $[A_0,B_0,C_0]$.
In addition, the modular symmetry is manifest for $\widetilde{W}_{s,[F_0]}$. 
We may use $\widetilde{W}_{s,[F_0]}H_t(u)$ as a superpotential by multiplying another modular form of the weight $t$.
Then, we can discuss the modular invariant $G=K+\ln |W|^2$.

Here, we give a comment on applications of $\widetilde{W}_{s,[F_0]}$ as well as  $\widetilde{W}_{s,[F_0]}H_t(u)$.
Suppose that the flux-induced superpotential $W(u)$ has the minimum at $e^{2\pi i /3}$.
We add $\varepsilon\widetilde{W}_{s,[F_0]}$ or   $\varepsilon\widetilde{W}_{s,[F_0]}H_t(u)$ as a correction term, where the solution of $F_0=0$ is not $e^{2\pi i /3}$ and $|\varepsilon|\ll 1$.
Then, the minimum is deviated from $u=e^{2\pi i/3}$ to $u=e^{2\pi i /3} + \delta$ slightly.
Such a small deviation from the fixed point is useful to realize fermion mass hierarchies \cite{Feruglio:2021dte,Novichkov:2021evw,Petcov:2022fjf,Kikuchi:2023cap,Abe:2023ilq,Kikuchi:2023jap,Abe:2023qmr,Petcov:2023vws,Abe:2023dvr,deMedeirosVarzielas:2023crv,Kikuchi:2023fpl}.

Moreover, the function $f_{s,[F_0]}$, which is defined above, has an interesting property under the Gauss composition.
We can define the following transformation:
\begin{align}
    [F_1] : f_{s,[F_0]} \to f_{s,[F_1]*[F_0]}.
\end{align}
For example, for $D=-20$, there appears the following $\mathbb{Z}_2$ symmetry:
\begin{align}
    &[1,0,5]:f_{s,[1,0,5]}\to f_{s,[1,0,5]},\notag\\
    &[1,0,5]:f_{s,[2,2,3]}\to f_{s,[2,2,3]},\notag\\
    &[2,2,3]:f_{s,[1,0,5]}\to f_{s,[2,2,3]},\notag\\
    &[2,2,3]:f_{s,[2,2,3]}\to f_{s,[1,0,5]}.
\end{align}

\bibliography{references}{}
\bibliographystyle{JHEP} 
\end{document}